\documentclass[12pt]{article}
\usepackage[utf8]{inputenc}
\usepackage{amssymb,amsmath,amsthm}
\usepackage[english]{babel}
\usepackage[T1]{fontenc}
\usepackage[dvipsnames]{xcolor}
\usepackage{appendix} 
\usepackage{bm}
\usepackage{lipsum}
\usepackage{hyperref}
\usepackage{geometry}
\usepackage{graphicx}
\usepackage{siunitx}
\usepackage{bbm}
\hypersetup{colorlinks,linkcolor={blue},citecolor={blue},urlcolor={blue}} 
\usepackage[authoryear]{natbib}
\usepackage{gensymb}
\usepackage[normalem]{ulem}
\newcommand{\betab}{\mbox{\boldmath \(\beta\)}}
\newcommand{\lambdaV}{\mbox{\boldmath \(\lambda\)}}
\def\ss{\mathbf{s}}

\def\SS{\mathcal{S}}
\def\R{\mathbb{R}}
\def\pr{\text{P}}

\title{Bayesian space-time gap filling for inference on extreme hot-spots: an application to Red Sea surface temperatures}
\author{$^1$Daniela Castro-Camilo, $^2$Linda Mhalla, $^3$Thomas Opitz}

\footnotetext[1]{
\baselineskip=10pt School of Mathematics and Statistics, University of Glasgow, UK.}
\footnotetext[2]{ 
\baselineskip=10pt Department of Decision Sciences, HEC Montr\'eal, Canada}
\footnotetext[3]{ 
\baselineskip=10pt Biostatistics and Spatial Processes, INRAE, Avignon, France}

\date{\today}

\begin{document}

\maketitle

\baselineskip=16pt
\begin{center}
{\large{\bf Abstract}}
\end{center}
We develop a method for probabilistic prediction of extreme value hot-spots in a spatio-temporal framework, tailored to big datasets containing important gaps. In this setting, direct calculation of summaries from data, such as the minimum over a space-time domain, is not possible. 
To obtain predictive distributions for such cluster summaries, we propose a two-step approach. We first model marginal distributions with a focus on accurate modeling of the right tail and then, after transforming the data to a standard Gaussian scale, we estimate a Gaussian space-time dependence model defined locally in the time domain for the space-time subregions where we want to predict.
In the first step, we detrend the mean and standard deviation of the data and fit a spatially resolved generalized Pareto distribution to apply a correction of the upper tail. To ensure spatial smoothness of the estimated trends, we either pool data using nearest-neighbor techniques, or apply generalized additive regression modeling. To cope with high space-time resolution of data, the local Gaussian models use a Markov representation of the Matérn correlation function based on the stochastic partial differential equations (SPDE) approach. In the second step, they are fitted in a Bayesian framework through the integrated nested Laplace approximation implemented in \texttt{R}-INLA. Finally, posterior samples are generated to provide statistical inferences through Monte-Carlo estimation. Motivated by the 2019 Extreme Value Analysis data challenge, we illustrate our approach to predict the distribution of local space-time minima
in anomalies of Red Sea surface temperatures, using a gridded dataset ($11315$ days, $16703$ pixels) with artificially generated gaps. In particular, we show the improved performance of our two-step approach over a purely Gaussian model without tail transformations.



\par\vfill\noindent
{\bf Keywords:} extreme-value theory, generalised additive modeling, INLA, Red Sea, sea surface temperature, SPDE.\\

\baselineskip=20pt


\newpage
\section{Introduction}~\label{sec:introduction}
Large georeferenced datasets, often based on remote sensing or reanalysis of smaller-sized observation data, have become abundant in the domains of climate, environment, and ecology. However, the spatial and temporal resolution in such datasets may not be fine enough for certain purposes. Moreover, gaps may arise in such datasets when sensors are defective or cannot provide useful data, for instance, due to cloud occlusion or orbital characteristics when using satellite-based instruments.
In this setting of gap filling and prediction of partially observed data in the Earth and atmospheric sciences \citep[\emph{e.g.}, ][]{Yuan2011,Henn2013,Mariethoz2012,Wang2012,Henn2013,C2017,Xing2017,Yin2017,Padhee2019}, we aim to provide probabilistic predictions of summary statistics of the data process for a space-time domain, and in particular to focus on episodes with simultaneous extreme values of a continuous variable such as temperature. By taking into account the spatial and temporal dependence, we seek inference for summary functionals of space-time hot-spots, such as the minimum value of the process over a spatial region during a certain period of time. 

Our work was motivated by the 2019 Extreme Value Analysis (EVA) data challenge to which we participated with our team named \emph{BeatTheHeat}. Its goal was to predict the minimum of anomalies of sea surface temperature (SST) of the Red Sea over regions where data have been artificially masked, and predictions for $162,000$ space-time cylinders with a spatial radius of $50$km and duration of $7$ days had to be provided.  
The Red Sea is home to over $300$ species of coral and $700$ species of fish, some of them endemic~(\citealp{spalding2001world}; \href{https://www.fishbase.in/search.php}{www.fishbase.in}).
This rich ecosystem is being threatened by global warming.
The 2014 International Panel on Climate Change report indicates that the average SST  of the Indian Ocean has increased by $0.65^{\circ}$C over the period 1950-2009~\citep{HoeghGuldberg2014climate}. Summertime SSTs for the past decade have remained, on average, $1.46^{\circ}$C above the historical mean~\citep{cantin2010ocean}.
Coral reefs are particularly sensitive to modest increases in background seawater temperature; a continued increase of $1^{\circ}$C or more above the historical maximum SST can result in substantial coral damage~\citep{cantin2010ocean}.

There is a need to understand the spatio-temporal dynamics governing SST,  and in particular their extremes, in the Red Sea. In this work, we consider temperature anomalies that were constructed by subtracting a spatio-temporal mean from gridded daily SST observations provided by the Operational SST and Sea Ice Analysis (OSTIA; \citealt{donlon2012operational}), a satellite-based data system designed for numerical weather prediction. The data encompasses $31$ years (1985--2015) of daily simulations over $16,703$ locations in the Red Sea. 

Geostatistics provides a wide range of models to describe and predict spatio-temporal phenomena. Historical and current developments are usually centered around Gaussian processes, which have appealing theoretical and computational properties~\citep{cressie2015statistics}.
Extreme-value analysis provides more specific tools to model extremes in space and time, although they usually pose a greater computational challenge.
In this work, we combine both disciplines to predict high-temperature anomalies over space and time in the Red Sea. 

Due to the large size of the dataset, care is needed to develop inference procedures that scale well with large datasets. 
We propose to model the marginal effects and dependence structure separately following a two-step approach. First,  we fit a model for marginal distributions and use it to transform data to a common standard Gaussian scale. Then, stationary Gaussian process models are used locally in time to capture spatio-temporal dependence. 

To describe non-stationary marginal anomalies at each site in the first step, we start by removing long-term temporal trends at a yearly scale, assumed to be stationary over space, by correcting data using empirical means and variances. 
Subsequently, we assume  temporal stationarity, and we seek to improve the upper tail representation  through a model for data above a stationary threshold close to the global $75\%$-quantile. We estimate a tail model with space-varying parameters based on a Poisson regression for the rate of threshold exceedances, and on a generalized Pareto (GP) model for the excesses above the threshold. For these marginal models, we compare three approaches that allow generating relatively smooth parameter surfaces. First, we consider the data without marginal transformation as a benchmark to assess the added value of correcting the tail. Second, we rely on the generalized additive modeling (GAM) framework \citep{Wood2003,Wood2006} where we use the latitude, longitude, and distance to the coastline as covariates to describe the spatial pattern in the tails. This framework allows for flexible modeling of covariate effects and has previously been used in univariate and multivariate extreme-value settings to capture non-stationarity in environmental data \citep{Pauli2001, chavez_davison, Jonathan2014, Mhalla_DC_CD}. Third, we implement spatial pooling of data using a nearest-neighbor (NN) approach to infer the tail behavior at each location. 

SSTs in the Red Sea are known to change smoothly in space and time~\citep{chaidez2017decadal}.
Therefore, in the second step of our approach, we describe the space-time dependence between Gaussian-transformed SST anomalies by a spatio-temporal stationary Gaussian term characterized by a stationary Mat\'ern covariance structure in space and first-order temporal autoregression. 
For fast likelihood-based inference and probabilistic prediction with large gridded datasets, we use the stochastic partial differential equations (SPDE) approach. It provides a numerically convenient Gauss--Markov approximation to the Mat\'ern covariance~\citep{lindgren.al.2011}, and we combine it with the computationally-efficient integrated nested Laplace approximation (INLA; \citealp{rue2009approximate,Rue.al.2017,Opitz.2017b,Opitz.al.2018,castro2019spliced}) for Bayesian inference. 

A numerical comparison of the performances of our three models with different marginal pretransformations (none, GAM, NN) is made using the tail-weighted continuous ranked probability score (twCRPS) \citep{Gneiting2011}. Our models perform better than all other contributions to the EVA challenge, and the tail correction provides a significant improvement over models without marginal transformation. 

The remainder of the paper is organized as follows.~Section~\ref{sec:data} presents the data through exploratory analyses and graphical illustrations. Section~\ref{sec:modeling} describes our modeling approaches, for which results and prediction performances are discussed in Section~\ref{sec:results}. 
Conclusions and an outlook towards further improvements in our model are given in Section~\ref{sec:discussion}.

\section{Red Sea surface temperature data}~\label{sec:data}
We describe the main features of the SST anomalies and provide exploratory graphical support for our modeling choices. 

\subsection{Description of data and  space-time hotspots}

The Sea Ice Analysis (OSTIA) system produces daily SST data in the Red Sea at a spatial resolution of 1/20{\degree}. We consider the period between January 1985 and December 2015. We denote by $Y(\ss,t)$  the Red Sea surface temperature at location $\ss\in\SS\subset\R^2$ and time $t\in\mathcal{T} = \{1,\ldots,T\}$, where $|\SS| = 16703$ and $T=11315$. Our analysis focuses on anomalies $A(\ss,t)$ where
$$A(\ss,t) = Y(\ss,t) -m(\ss,t) ,$$
and $m$ is a space-time mean surface; see \cite{huser2020eva} for details on its estimation based on averages for each pixel-month combination. Anomalies remove recurrent local climatic trends and allow for easier interpretation of short-term variability and large-scale anomalous behavior with respect to what we observe on average at a given site and time of the year. 
Large-scale extreme temperatures are likely to negatively affect the environmental and ecological systems in the Red Sea, and we are interested in measuring the impact of extreme anomalies occurring jointly in a specific space-time region.

While the original dataset does not contain any gaps with missing data, such gaps have been inserted artificially in the dataset to compare predictive approaches; see \cite{huser2020eva}. During each month, the data for a randomly selected subdomain of the Red Sea were removed. Using the remaining observations, our goal is to predict a summary functional of space-time subregions contained in these gaps, \emph{i.e.},  where no observations are available. 
Specifically, for each pair $(\ss,t)\in\mathcal{X}=\R^2\times\mathcal{T}$, we consider the following cylindrical neighborhood $$ \mathcal{N}(\ss,t) = \mathcal{B}(\ss,r)\times\{t-3,t-2,t-1,t,t+1,t+2,t+3\}\cap\mathcal{X},$$
where $\mathcal{B}(\ss,r)$ is a ball centered at $\ss$ with radius $r=50$km. We are interested in predicting the distribution of summary functionals  describing  \emph{extreme space-time hot-spots}, by using the minimum summary: 
$$X(\ss, t) = \min_{(\ss',t')\in\mathcal{N}(\ss,t)}A(\ss',t').$$
For the EVA competition, the validation  dataset of observations to be predicted, $X(\ss_i,t_i)$, $i=1,\ldots, 162,000$, consists of $324$ distinct time values $t_i$, and $500$ points $\ss_i$ for each of these time points. Our aim is to successfully predict these summaries  with a focus on extreme quantiles of $X(\ss,t)$, which characterize extreme hot-spots of SST anomalies. 

The Red Sea spans a relatively large region on the globe, such that it is not straightforward to choose a sound $2$D coordinate system for models that take into account spatial distances. While 1 degree of latitude corresponds to $111.2$km, we find that 1 degree of longitude corresponds to $96.3$km at the northern limit and to $108.8$km at the southern limit of the Red Sea. We here switch to a metric system with km unit to avoid artificial anisotropy effects as best as possible, by multiplying latitude degree values with $111.2$ and longitude degree values with the average between $96.3$ and $108.8$.

\subsection{Exploratory analysis \label{subsec:explo}}
\begin{figure}[!t]
    \centering
    \includegraphics[scale=0.25]{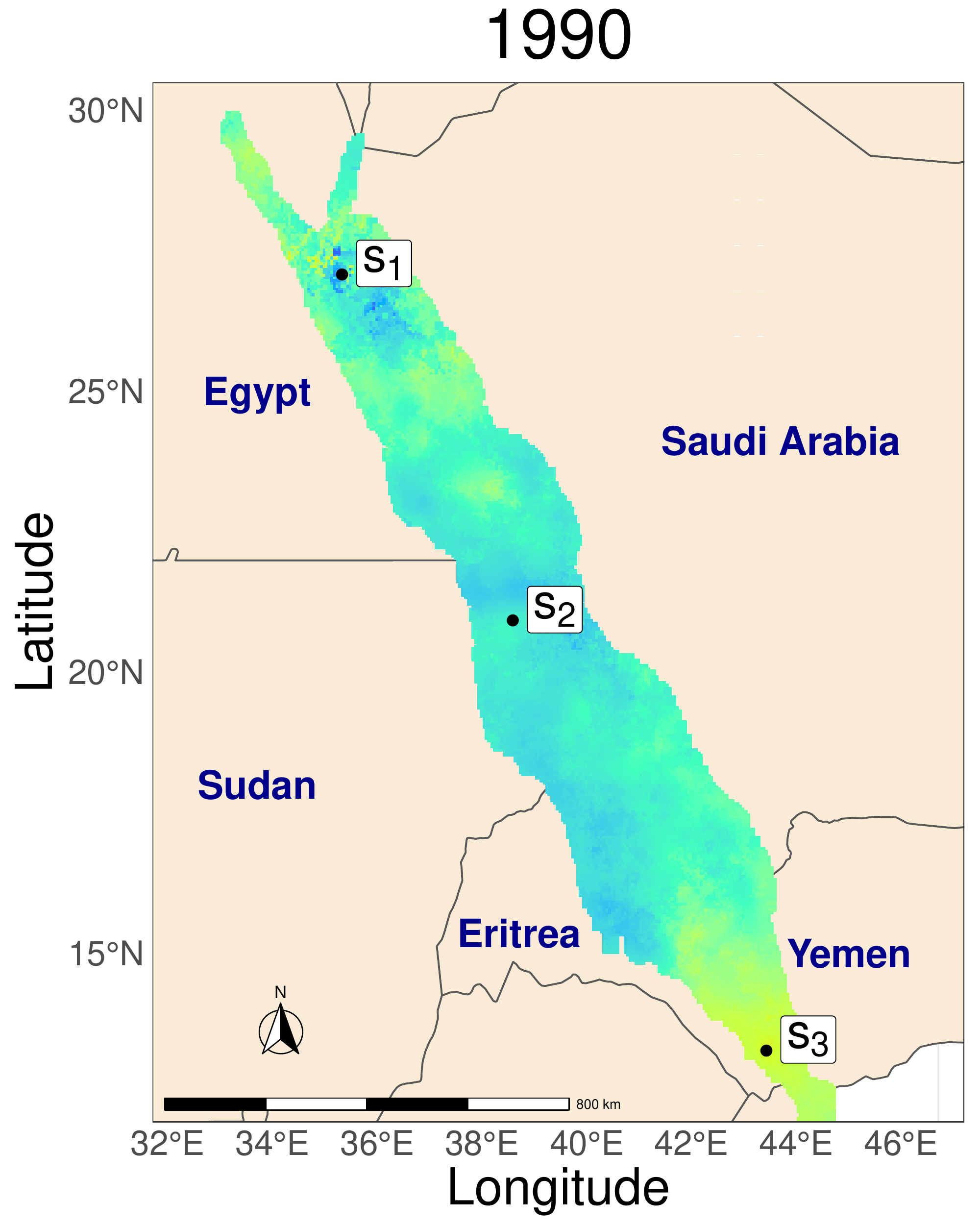}
    \includegraphics[scale=0.25]{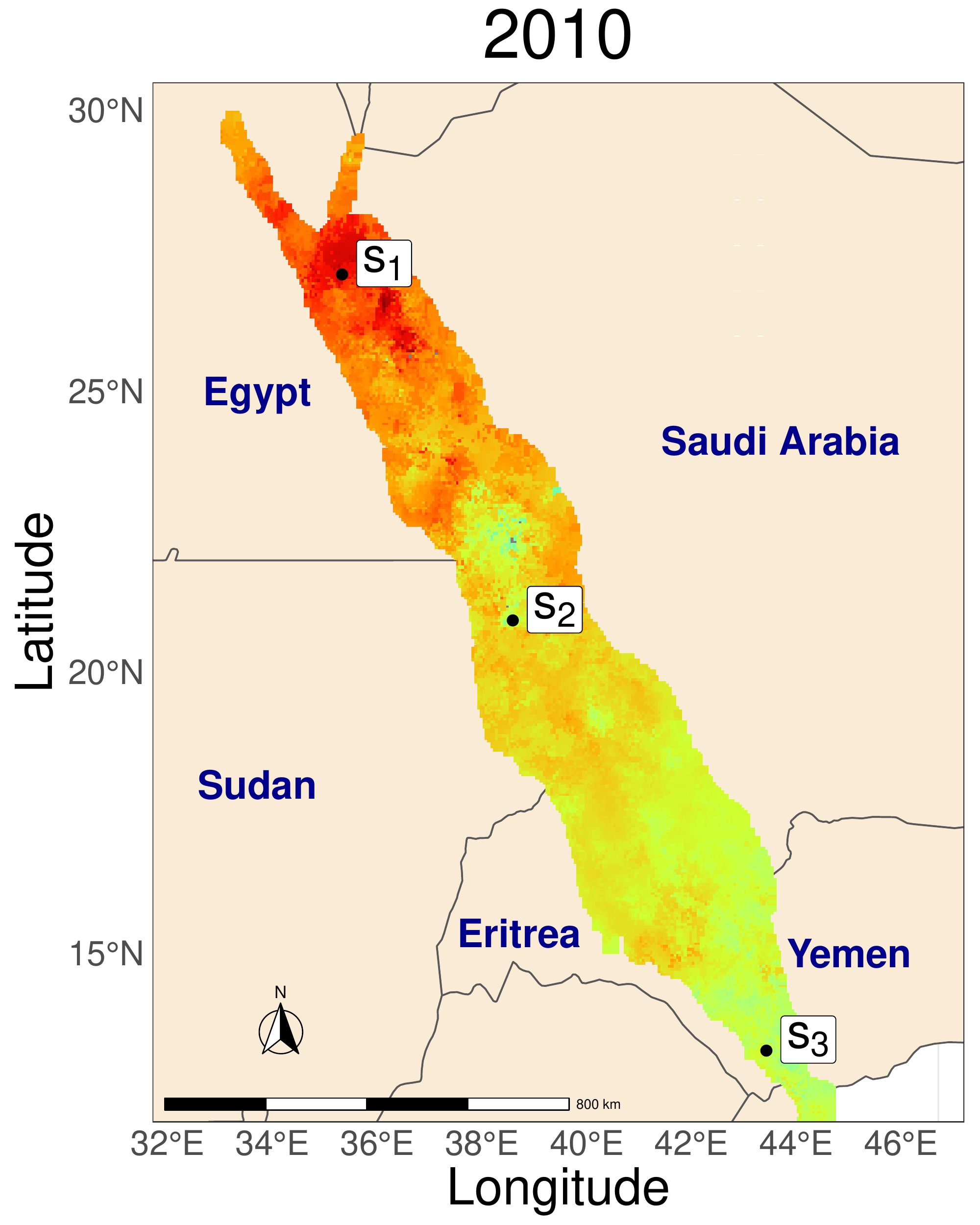}
    \includegraphics[scale=0.25]{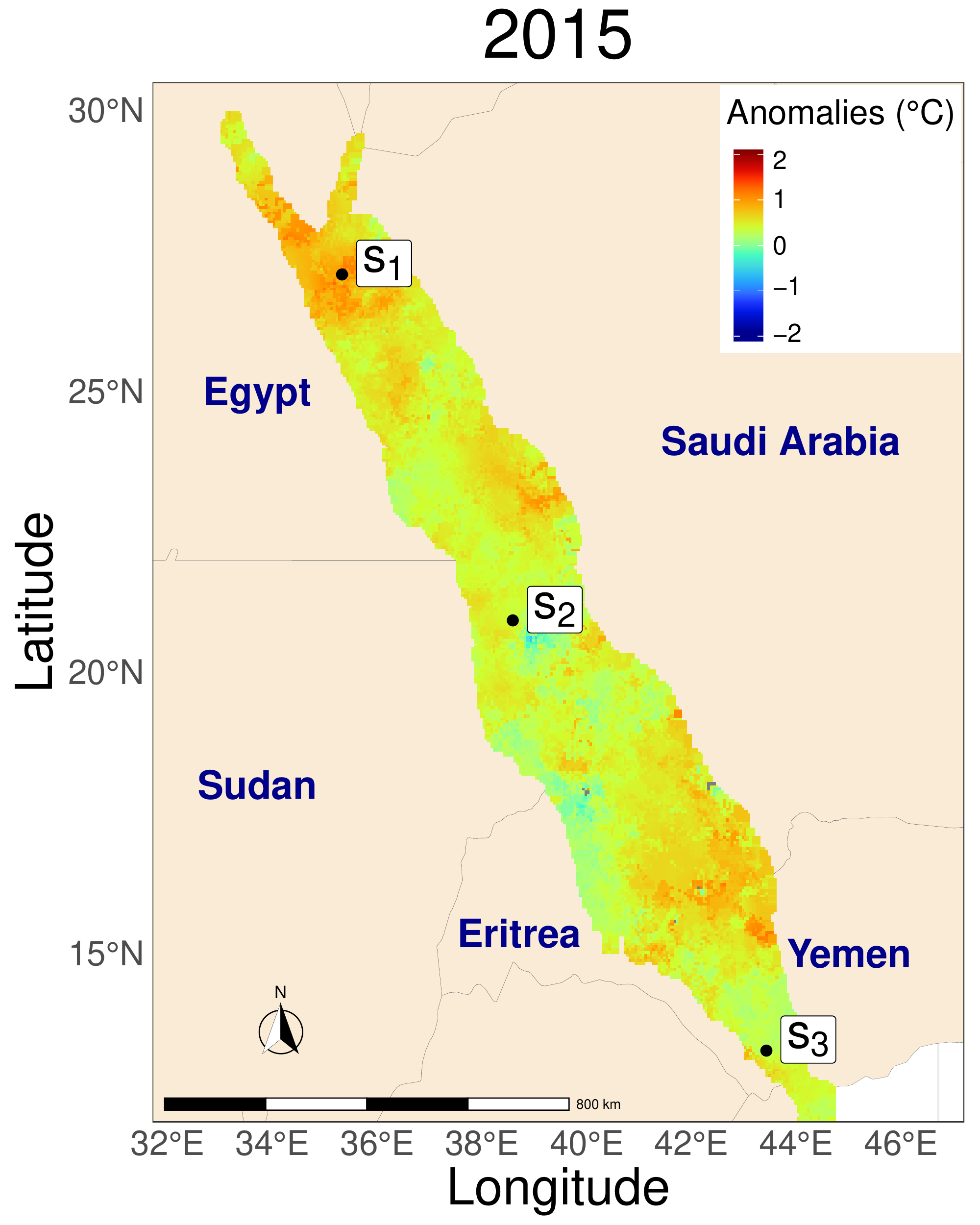}
    \caption{Annual average anomalies for three different years. Data at the three locations indicated on the maps are used in Figure~\ref{fig:qqplots}.}
    \label{fig:AnnualAvgYears.pdf}
\end{figure}

Despite the preliminary removal of mean trends according to location and month, strong nonstationarity remains in the data $A(s,t)$, particularly in the upper tail. To illustrate this, Figure~\ref{fig:AnnualAvgYears.pdf} shows maps of annual average anomalies in the Red Sea for three different years using the same colour scale, highlighting the fact that average anomalies vary strongly between years. We can also see a mild spatial effect in the 2010 map, but overall, the temporal effect seems to drive the nonstationarity observed in the average anomalies.


\begin{figure}[!t]
    \centering
    \includegraphics[width=0.45\linewidth]{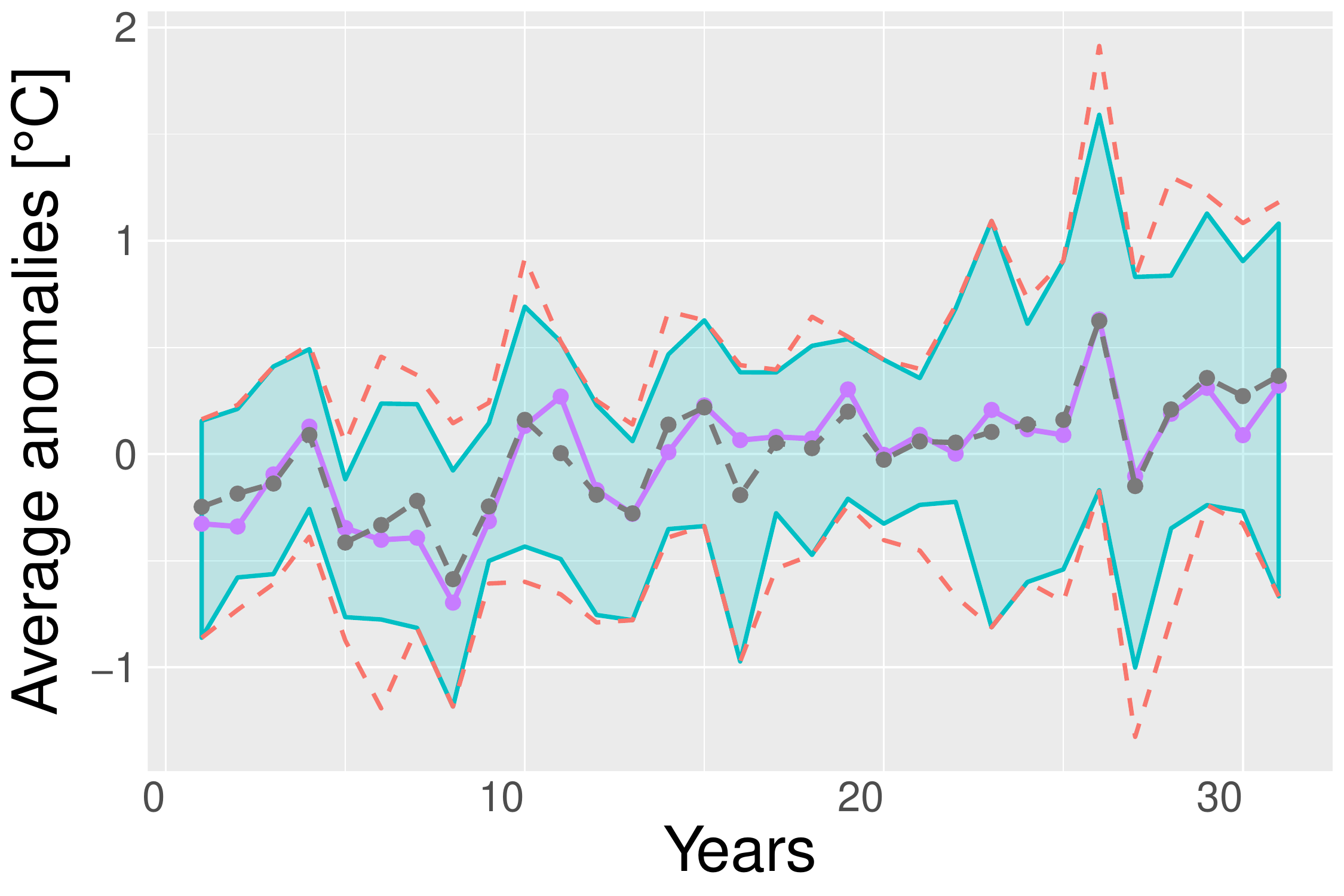}
    \includegraphics[width=0.45\linewidth]{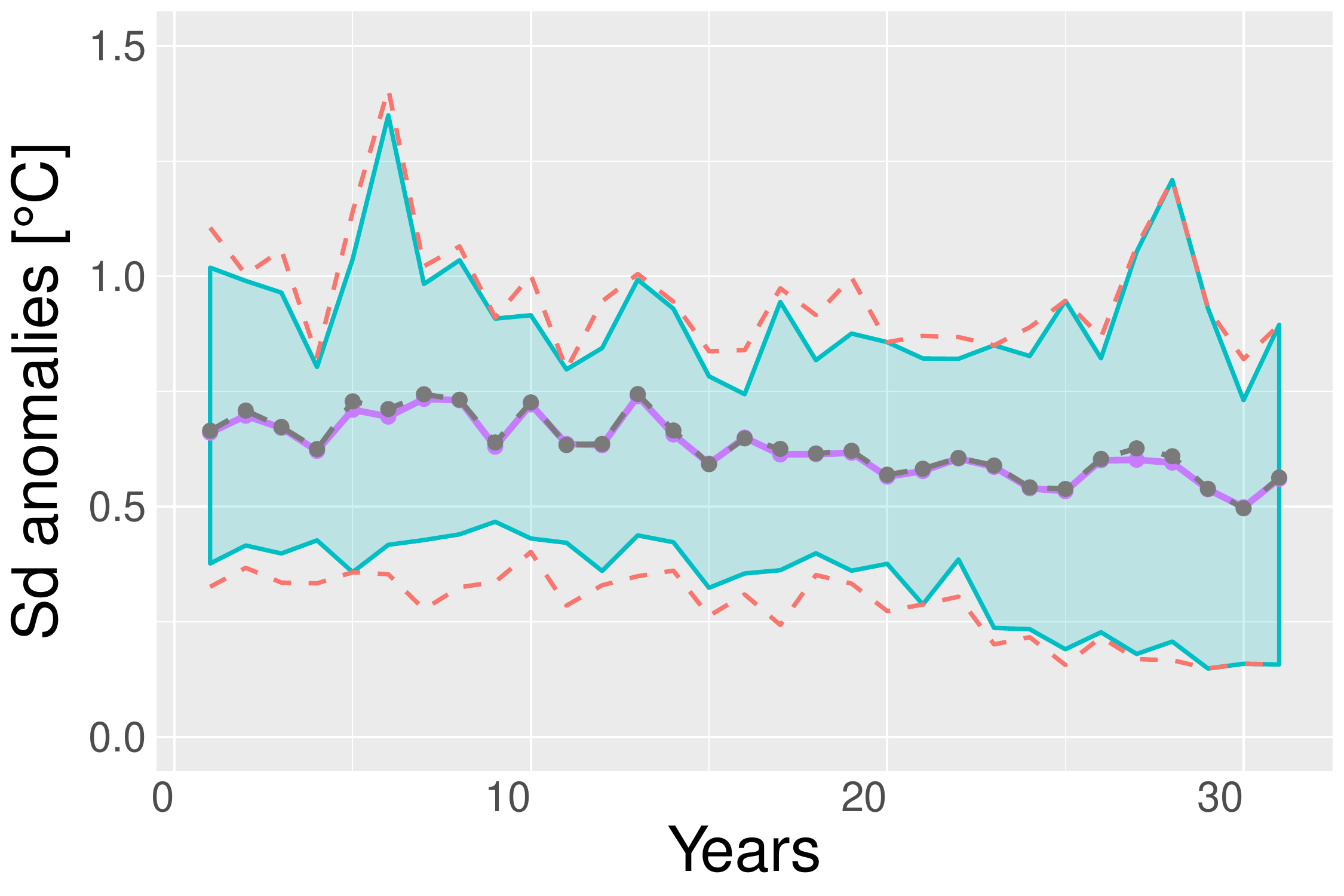}
    \caption{Annual average anomalies (left) and standard deviations (right). Functional boxplots summarise the behavior of the statistics across all locations: the blue polygons contain the 50\% central curves, while the outer coral dashed lines correspond to the minimum and maximum curves. The solid purple lines correspond to the overall statistics by year. The gray dashed lines correspond to the fitted mean and standard deviation as specified by the regression model~\eqref{eq:gam_Gauss}.}
    \label{fig:AnomaliesYear.pdf}
\end{figure}

To further illustrate the behavior of SST anomalies through years, we compute the annual average anomalies along with their standard deviations for every location in each year. These location and scale statistics can be treated as functional data over the years (one function per location), and displayed using the functional boxplots \citep{sun2011functional} in Figure~\ref{fig:AnomaliesYear.pdf}. To highlight the effect of the years at each location, we also plot the overall statistics by year (\emph{i.e.}, average and standard deviation over all locations for every fixed year). We can see that there is an upward annual trend over the averages, and a mild downward annual trend affecting the location-wise standard deviations of the SST anomalies. 

\begin{figure}[!t]
    \centering
    \includegraphics[width=0.35\linewidth]{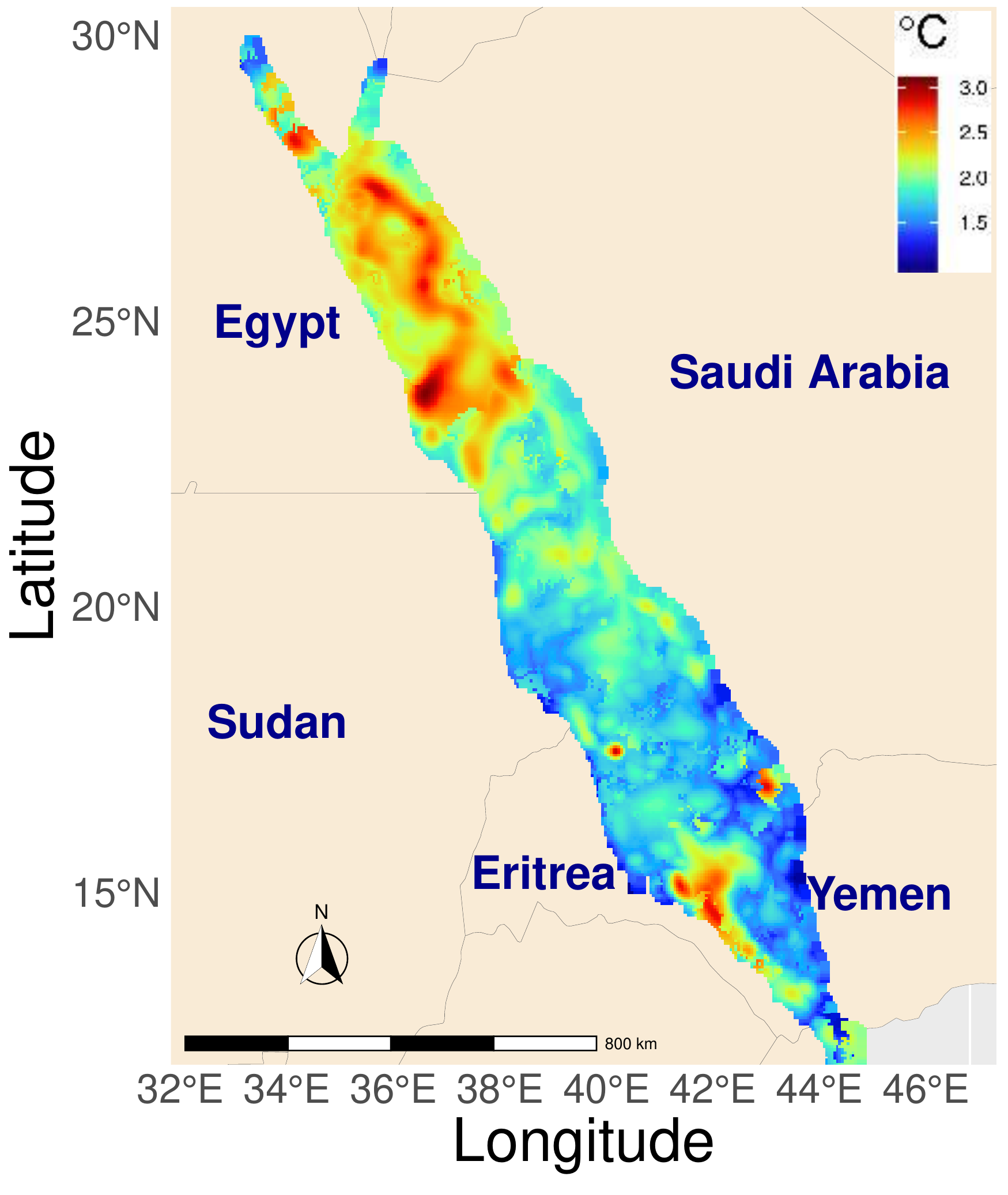}
    \includegraphics[width=0.35\linewidth]{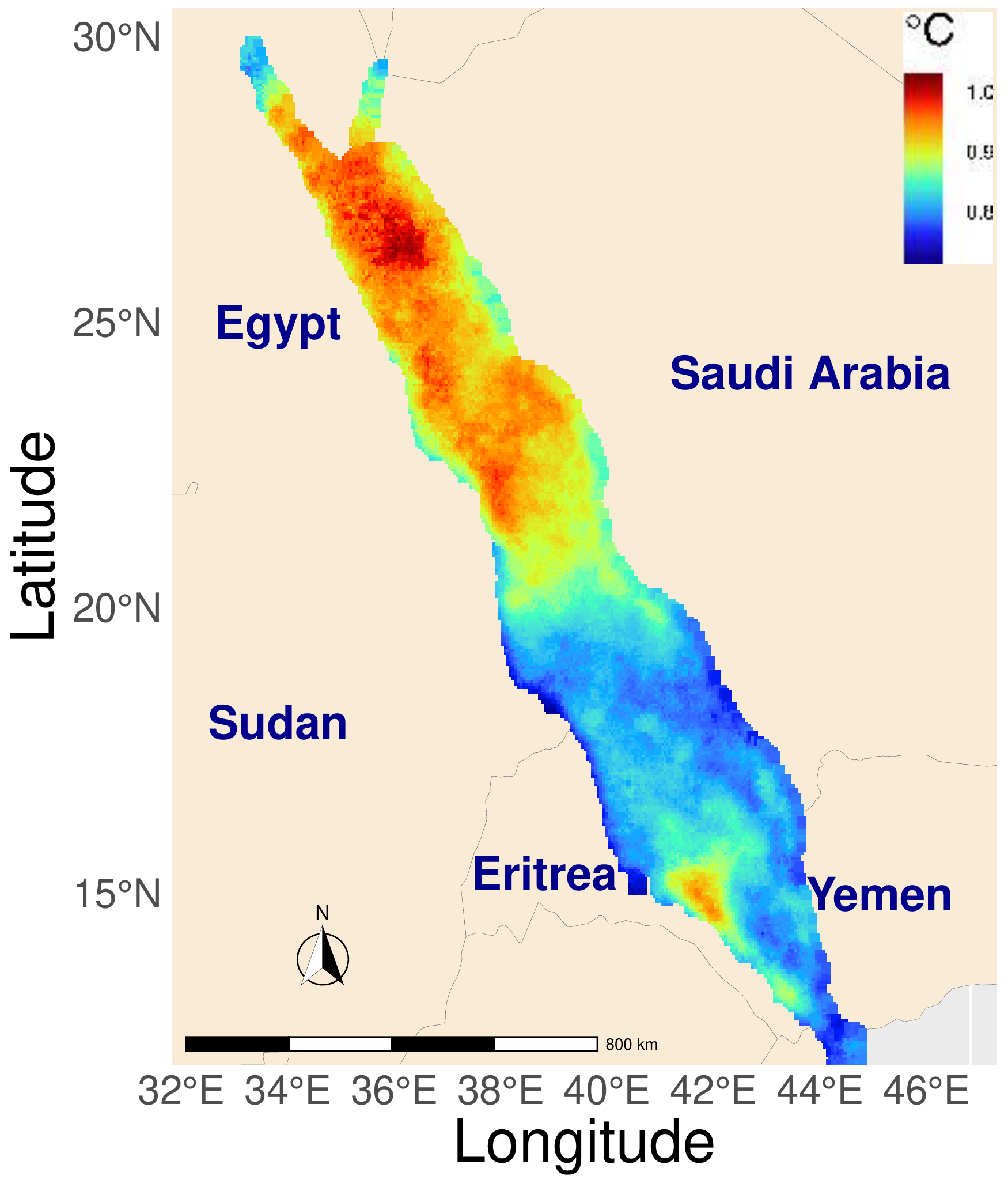}
    \caption{Location-wise empirical range (left) and mean exceedance value (right).}
    \label{fig:TailAnomaliesLocations.pdf}
\end{figure}

To unveil spatial nonstationarity in the tail of the SST anomalies, we characterize the distribution of threshold exceedances defined as those SST anomalies exceeding $0.49$, corresponding to an overall exceedance probability of approximately $20\%$, and used later for tail modeling. 
Figure~\ref{fig:TailAnomaliesLocations.pdf} shows the location-wise empirical range of exceedances (\emph{i.e.,} the difference between the maximum and minimum values) as well as the mean exceedance value, $\mathbb{E}\{A(\ss,t)\mid A(\ss,t) >u\}$. We can see a clear North to South pattern in both maps, consistent with a strong spatial effect in the upper tail of the SST anomalies.

\begin{figure}[!t]
    \centering  
    \includegraphics[width=.9\linewidth]{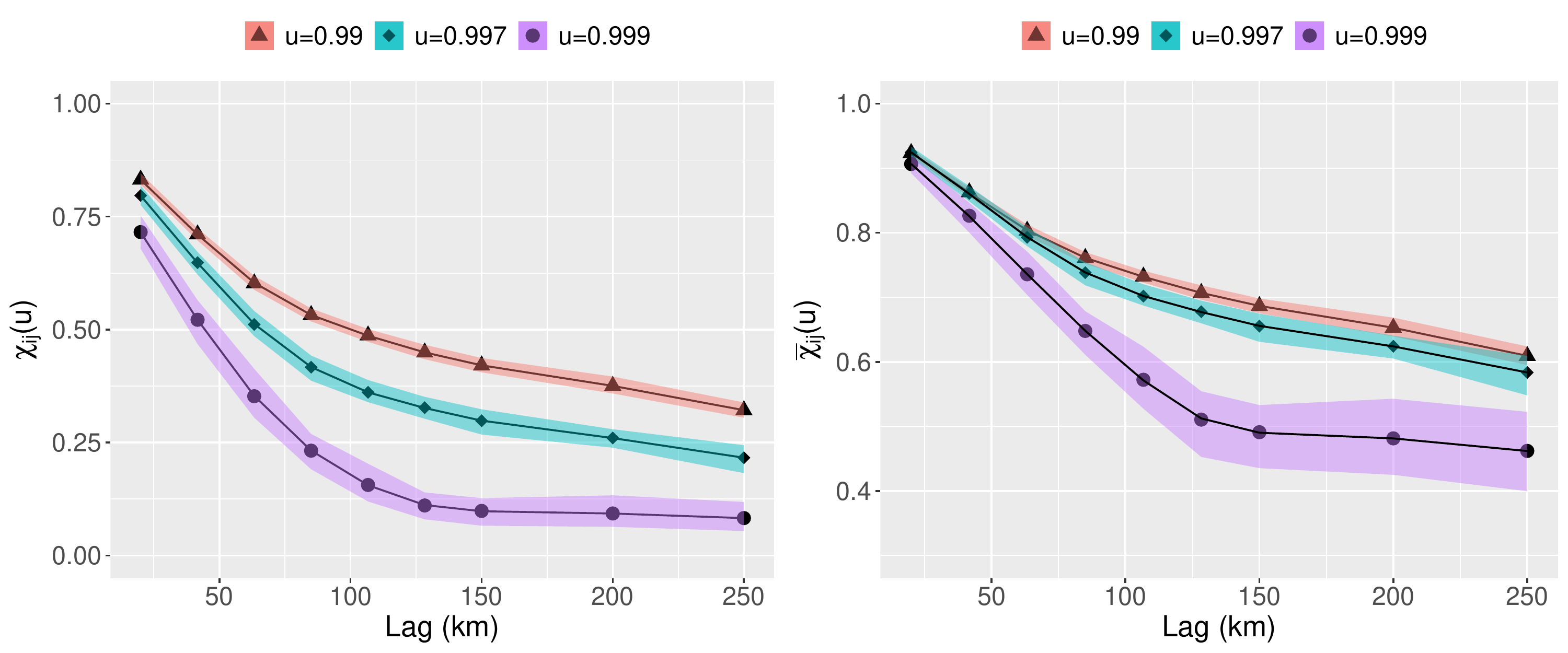}
    \caption{Empirical estimates of $\chi_{ij}(u)$ (left) and $\bar{\chi}_{ij}(u)$ (right) for the SST anomalies data. The filled regions represent the $95\%$ bootstrap confidence bands.}
    \label{fig:chi_chibar}
\end{figure}

We conclude this section by studying bivariate summaries of extremal dependence. Specifically, we analyze two well-know pairwise measures of tail dependence based on conditional exceedance probabilities, namely
\begin{align*}
    \chi_{ij}(u) &= \pr \left[ F_i\{A(\ss_i,\cdot)\}>u\mid F_j\{A(\ss_j,\cdot)\}>u \right],\\
    \bar{\chi}_{ij}(u) &= \frac{2\log(\pr[ F_i\{A(\ss_i,\cdot)\} >u])}{\log(\pr[F_i\{A(\ss_i,\cdot)\}>u, F_j\{A(\ss_j,\cdot)\} >u])}-1,
\end{align*}
where $F_i$ is the distribution function at location $i$ and $A(\ss_i,\cdot)$ is the time series at location $i$.
Together, $\chi=\lim_{u\to1} \chi_{ij}(u)$ and $\bar{\chi}= \lim_{u\to1} \bar{\chi}_{ij}(u)$ summarise the strength of the spatial tail dependence where $\chi>0$ characterizes \emph{asymptotic dependence}, while $\bar{\chi}<1$ may arise if in the case of \emph{asymptotic independence} ($\chi=0$). Figure~\ref{fig:chi_chibar} displays bootstrap estimates and confidence bands for $\chi_{ij}(u)$ and $\bar{\chi}_{ij}(u)$, where we fix $u$ at three high probabilities $u=0.99,0.997,0.999$ and consider pairs of locations $(i,j)$ separated by different spatial lags. For computational reasons and due to the strong spatial dependence in data, the estimation is conducted on a subset of the available locations, obtained by regular sampling of 1558 locations.
The bootstrap procedure relies on $300$ bootstrap samples drawn by resampling from the set of pairs of locations. The resulting estimates of $\chi_{ij}(u)$ point towards asymptotic independence as they strongly decrease with larger spatial lags and higher probabilities. The estimates of $\bar{\chi}_{ij}(u)$ remain clearly below $1$ for all positive distances, which confirms asymptotic independence in data. They show significantly lower values for higher probabilities, which indicates particularly fast joint tail decay rates. Moreover, this analysis has been done without removing temporal trends in margins, such that even stronger evidence of relatively weak asymptotic independence can be expected after removing such trends. 

\section{Space-time predictive modeling}
\label{sec:modeling}
In this section, we describe our two-step approach to model trends in marginal distributions and local space-time dependence. 
\subsection{Marginal modeling}
We consider three approaches to pretransforming data before fitting local Gaussian models. 
\subsubsection{Model 1: no marginal transformations}
\label{subsec:notransform}
As a reference model allowing us to assess the utility of marginal pretransformations, we implement an approach where the local Gaussian dependence model is fitted to raw anomalies $A(s,t)$ without preliminary marginal transformations. 

\subsubsection{Model 2: generalized additive regression}\label{subsec:marginalmodeling}
As discussed in Section~\ref{subsec:explo}, the SST anomalies exhibit 
a residual temporal trend in the mean and a spatial non-stationary pattern in the tails. We first remove the long-term time trends observed in Figure~\ref{fig:AnomaliesYear.pdf} by fitting a Gaussian location-scale 
regression model. Recall that we denote by $A(\ss,t)$ the SST anomaly at location $\ss \in \mathcal{S}$ and time $t \in \mathcal{T}$. Then, we assume the following model structure $A(\ss,t)= \mu(t)+e(t)$, where $e(t)\sim\mathcal{N}(0,\sigma^2(t))$ and
\begin{eqnarray}
\mu(t) &=& \beta_{\mu,i} \mathbbm{1}_{\texttt{yod}_t=i}, \nonumber\\
\log\{\sigma(t) \} &=& \beta_{\sigma,i} \mathbbm{1}_{\texttt{yod}_t=i}, \quad i=1,\ldots, 31 \label{eq:gam_Gauss}
\end{eqnarray}
where $\texttt{yod}_t$ denotes the year of day $t$ and $\beta_{\mu,i}$ and $\beta_{\sigma,i}$ are the average mean and standard deviation of the year $1984+i$. Owing to the large number of observations in the dataset as well as the spatial pooling of the data, this simple model specification should be able to capture the observed temporal non-stationarity while maintaining the inherent spatio-temporal structure of the large anomalies that we ought to predict.

To reduce the computational cost of fitting the model in~\eqref{eq:gam_Gauss}, we use a sub-sampling strategy by randomly selecting one site out of 50 and keeping all the observations over time. Our approach is sensible since original SST anomalies were obtained using a spatial multi-scale optimal interpolation scheme~\citep{donlon2012operational}. 

As the interest of this work is in the prediction of jointly large anomalies, we propose to improve the tail modeling of the standardized anomalies. More precisely, the SST anomalies $A(\ss,t)$ are normalized, in a first step, to a standard Gaussian scale using the fitted model~\eqref{eq:gam_Gauss}, \emph{i.e.},
\begin{equation}
    \tilde{Z}(\ss,t) = \dfrac{A(\ss,t) - \hat{\mu}(t)}{\hat{\sigma}(t)}, \quad \ss \in \mathcal{S}, \ t \in \mathcal{T}, \label{eq:transformed_data}
\end{equation}
where $\tilde{Z}(\ss,t)$ are now supposed (approximately) stationary over time. Then, in a second step, we refine the model in the tails by considering a spatial Generalized Pareto (GP) model for the exceedances of a high threshold $u$. Specifically, we assume a generalized additive model (GAM) where the scale and shape GP parameters are described by
\begin{eqnarray}
\log\{\sigma^{\text{GAM}}_{GP}(\ss)\} &=&  f_1(\texttt{lat}_\ss,\texttt{lon}_\ss) + f_2(\texttt{dist}_\ss),\nonumber\\
\xi^{\text{GAM}}(\ss) &=&  f_3(\texttt{lat}_\ss,\texttt{lon}_\ss) + f_4(\texttt{dist}_\ss),\label{eq:gp_gam}
\end{eqnarray}
where $\texttt{lat}_\ss,\texttt{lon}_\ss$ are the latitude and longitude of the site $\ss$, while $\texttt{dist}_\ss$ is its distance to the coast (in meters). Within the GAM framework, the unknown functions $f_2$ and $f_4$ are represented using reduced rank smoothing splines, and more specifically, the reduced rank isotropic thin plate splines \citep[Chapter 4]{Wood_book2017}. The smooth interaction terms $f_1$ and $f_3$ are constructed using the tensor product construction where the main effects of $\texttt{lat}$ and $\texttt{lon}$ are excluded \citep{Wood2006}. Using their respective fixed basis expansions, estimation of the unknown smooth functions boils down to inference on the coefficients of their model matrices. In other words, each smooth term $\mathbf{f}_j=(f_j(x_{j1}),\ldots,f_j(x_{jn}))^\top$ can be written as $\mathbf{f}_j= \mathbf{X}_j \betab_{j}$ where $\mathbf{X}_{j}$ is the model matrix constructed by evaluating the (known) basis functions at the observed values of the covariates, and $\betab_j$ the vector of its corresponding coefficients to be estimated. To avoid overfitting issues, each smooth function $f_j$ is associated with a smoothing penalty term reflecting its curvature and defined as $\betab^\top \mathbf{S}_j \betab$, where $\mathbf{S}_j$ is a known positive definite matrix that depends on the basis functions. In the special setting of tensor products, the wiggliness of the smooth surface is controlled by two smoothing parameters associated with each marginal direction. Then, given a set of smoothing parameters $\lambdaV=(\lambda_1, \ldots, \lambda_6)$, the estimated model coefficients are obtained by maximum penalized likelihood
\begin{equation}
    \hat{\betab} = \underset{\betab}{\mathrm{argmax}} \ \left\lbrace \ell(\betab) - \sum_{j=1}^6 \lambda_j \betab^{\top}\mathbf{S}_j \betab /2 \right\rbrace, \label{eq:beta_hat_gam}
\end{equation}
where $\ell(\betab)= \log f(\tilde{\mathbf{Z}} \mid f_1, f_2) = \log f(\tilde{\mathbf{Z}} \mid \betab)$ is the Generalized Pareto (GP) log-likelihood 
\begin{equation}
    f\{\tilde{Z}(\ss,t) \mid \betab\} = \left\{
    \begin{array}{ll}
        \dfrac{1}{\sigma^{\text{GAM}}_{GP}(\ss)} \left\lbrace 1 + \xi^{\text{GAM}}(\ss) \dfrac{\tilde{Z}(\ss,t) - u}{\sigma^{\text{GAM}}_{GP}(\ss)} \right\rbrace^{-1/\xi^{\text{GAM}}(\ss) -1} & \mbox{if } \xi^{\text{GAM}}(\ss) \neq 0, \\
        \dfrac{1}{\sigma^{\text{GAM}}_{GP}(\ss)} \exp \left\lbrace -\dfrac{\tilde{Z}(\ss,t) - u}{\sigma^{\text{GAM}}_{GP}(\ss)} \right\rbrace  & \mbox{if } \xi^{\text{GAM}}(\ss) = 0, \notag
    \end{array}
\right.
\end{equation}
and $\tilde{\mathbf{Z}}= \{ \tilde{Z}(\ss,t), \ss \in \mathcal{S}, \ t \in \mathcal{T}\}$. The smoothing parameters are estimated iteratively using the restricted marginal likelihood approach \citep{Wood2010,Wood2016}. Precisely, for an intermediate estimate $\hat{\betab}$ obtained through a Newton step based on \eqref{eq:beta_hat_gam} and for a fixed $\lambdaV$, \cite{Wood2016} propose to undertake a Bayesian view where $\hat{\betab}$ is a posterior mode for $\betab$ and the estimate of the smoothing parameters $\lambdaV$ is updated by maximizing a Laplace approximation of the log marginal likelihood of the GP model. Uncertainty measures for the model coefficient estimates include thus the uncertainty associated with the smoothing parameter estimation; see \citet[Section~4]{Wood2016} for details. This fitting procedure is implemented in the package \texttt{mgcv} \citep{Wood_book2017} available in the \texttt{R} statistical software, and its application in an extreme-value setting is illustrated in \cite{Youngman2019}.

Using an \emph{ad-hoc} trial and error procedure, we choose a fixed threshold at the value $u=0.75$ for all sites and all time points. This value corresponds roughly to the 78.3\% empirical quantile of the transformed data in~\eqref{eq:transformed_data}. To cope with the computational demands of fitting the GAM model in~\eqref{eq:gp_gam} to all threshold exceedances, we keep all the sites but randomly select one excess out of ten. 

Finally, to compensate for the choice of a stationary threshold, we characterize the spatial behavior of the frequency of extreme SST anomalies by modeling the exceedance probability $p^{\text{GAM}}(\ss) = \Pr\{ Z(\ss,t)> 0.75 \}$. More specifically, we model the number of exceedances using a Poisson regression model with a spatially varying intensity following the same specification as in~\eqref{eq:gp_gam}.

\subsubsection{Model 3: Spatial nearest-neighbour modeling}\label{subsec:marginallocalmodeling}
To compare the modeling of the tail behavior with smooth covariate effects described in Section~\ref{subsec:marginalmodeling}, we propose an alternative local modeling approach. It exploits the large number of observations and captures the spatial pattern in the large anomalies of the normalized observations $\tilde{Z}(\ss,t)$ by modeling tail non-stationarity separately for each pixel. Therefore,we pool together the values $\tilde{Z}(\ss,t)$ of the $40$ nearest neighbor pixels and of the pixel itself. Based on this sample,  we then calculate the pixel-specific likelihood estimates of the exceedance probability $p^{\text{NN}}(\ss)$ and of the generalized Pareto parameters  $\sigma^{\text{NN}}_{GP}(\ss)$ and $\xi^{\text{NN}}(\ss)$.

\subsection{Local dependence modeling}
\label{sec:local}
For predicting nonlinear cluster functionals such as minima or maxima, it is important to provide an appropriate model for all multivariate distributions of the space-time process. In particular, relatively fast and robust geostatistical kriging approaches using only the mean and the variogram \citep{cressie2015statistics}  are not suitable.
Moreover, the estimation of a dependence model for the full dataset may be impeded by the large size of the dataset, as in the case of the Red Sea data with more than $150$ million observations. Instead, we propose to use a local dependence model for subperiods, including the space-time sets over which we aim to predict cluster functionals.  We propose using a stationary Gaussian space-time dependence model for such subperiods, applied either directly to the anomalies $A(s,t)$, or after using a marginal model to transform margins to the standard Gaussian distribution. We then combine a latent variable approach with Gauss-Markov dependence structures in a Bayesian framework to cope with the high space-time resolution and to provide probabilistic predictions for cluster functionals.

As shown by Figure~\ref{fig:chi_chibar}, there is strong evidence in favor of asymptotic independence for the Red Sea SST data. 
Therefore, classical limit models of dependence from extreme-value theory \citep[max-stable processes or generalized Pareto processes, see][]{Davison.Padoan.Ribatet.2012,Ferreira.deHaan.2014,Thibaud.Opitz.2016} would be inappropriate,  in addition to being not easily tractable for prediction with large datasets. Gaussian dependence has already proved to be flexible for tails in some cases \citep{Bortot.Tawn.2000}.

A useful choice for modeling spatio-temporal dependence is to link spatial fields through temporal auto-correlation, 
and in particular first-order autoregression, which leads to sparse spatio-temporal precision matrices (\emph{i.e.}, inverse covariance matrices) in case of sparse spatial precision matrices. 
We consider the following stationary space-time process defined for discrete time with regular time steps:
\begin{align}\label{eq:spgaussian}
W(\ss,1) &= w_0 + \varepsilon_1(\ss), \nonumber\\
W(\ss,t+1) &= (1-\rho)w_0 + \rho W(\ss,t) + \sqrt{1-\rho^2}\varepsilon_{t+1}(\ss), \quad t=1,2,\ldots, t_{\max},
\end{align}
where $\varepsilon_t$, $t=1,2,\ldots, t_{\max}$ are centered Gaussian random fields with Mat\'ern covariance and $w_0\in\mathbb{R}$ is the mean parameter. The space-time precision matrix for the Cartesian product of a collection of sites and times corresponds to the Kronecker product of the corresponding purely spatial (Mat\'ern) and purely temporal (AR(1)) precision matrices. In our local models, we use $t_{\max}=9$, where we have added one day before and after the periods of seven days length, over which we seek to predict cluster functionals such as minima. Using larger temporal buffers around the prediction period would be possible and could further improve the prediction performance in cases where the conditional independence of $W(\ss,t+1)$ and $W(\ss,t-1)$ given $W(\ss,t)$, as suggested by first-order autoregression, does not hold. For reasons of computational cost, we do not explore this possibility in our data application.

Using the stochastic partial differential equation (SPDE) approach developed by \citet{lindgren.al.2011}, we can work with numerically convenient sparse precision matrices  that approximate the Mat\'ern covariance function. We shortly recall this framework and refer to \citet{Krainski.al.2018} for details.
The SPDE is given by
\begin{equation}
\label{eq:spde}
\left(\kappa^2 - \Delta \right)^{\alpha/2} \tau W(\ss)  = E(\ss),
\quad  \alpha=\nu+D/2, \quad \ss \in \mathbb{R}^d, 
\end{equation}
with the Laplace operator $\Delta y=\sum_{j=1}^D \partial^2 y/\partial^2 x_j$, and a standard Gaussian white noise process $E(\ss)$. Its unique stationary solution is a zero-mean Gaussian process $W(\ss)$ with Mat\'ern covariance function with shape $\nu>0$, fixed to $\nu=1$ in the following.  The marginal
variance is $\sigma_W^2=\Gamma(\nu)/\{\Gamma(\nu+D/2)(4\pi)^{D/2}\kappa^{2\nu}\tau^2\}$,
and the \emph{empirical range} parameter is $\sqrt{8\nu}/\kappa^2$, indicating the distance where we observe a correlation of approximately $0.1$. 
With a finite study region, appropriate boundary conditions ensure a unique solution, and we can set the boundary relatively far from the actual study region to render boundary effects negligible within the study area. 

\begin{figure}[!t]
    \centering
    \includegraphics[width=0.4\linewidth]{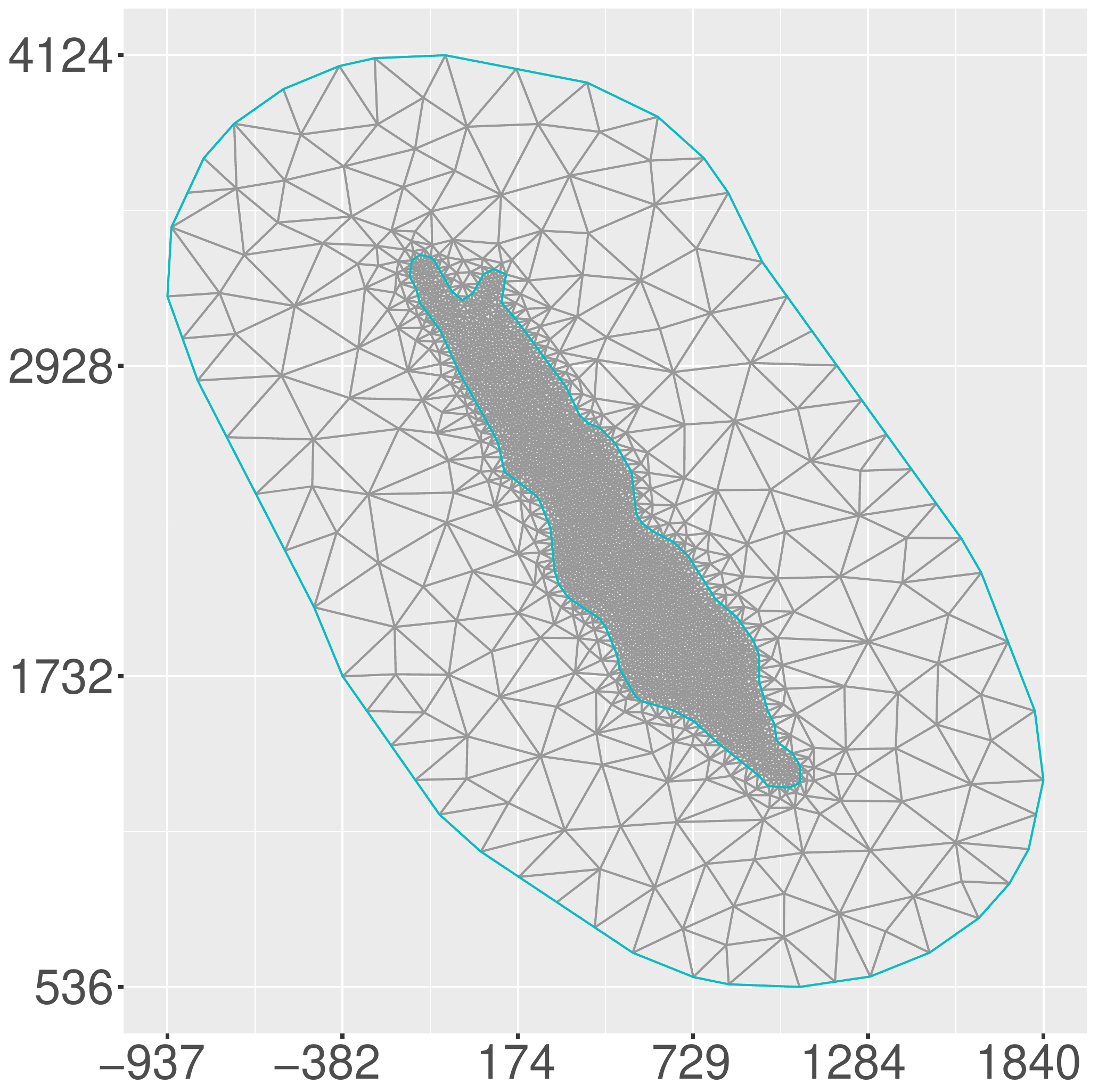}

    \caption{Triangulation mesh (km unit) used to discretize the study region for latent variable modeling with Gaussian variables located at the mesh nodes. The inner blue boundary delimits the study area, while the extension to the outer blue boundary ensures approximately stationary behavior of the SPDE solution within the study area.}
    \label{fig:mesh}
\end{figure}

The approximate Gauss--Markov solution to \eqref{eq:spde}
is given by constructing a discretization of space through a triangulation mesh, and by using  the representation $W(\ss) = \sum_{j=1}^{m_W} \omega_j \Psi_j(\ss)$ with ``pyramid-shaped" basis functions $\Psi_j(\ss)$, one for each mesh node. 
By solving the SPDE in the subspace spanned by the linear combination $W(\ss)$, we obtain $\bm \omega = (\omega_1,\ldots, \omega_{m_W})^T\sim \mathcal{N}(0,Q_W)$ with explicitly known precision matrix $Q_W$. The mesh used for the Red Sea in our data application is shown in Figure~\ref{fig:mesh}. It contains $m_W=2746$ nodes, and we choose a fine resolution in the study area (within the inner blue boundary), while a coarser mesh is used in the extension area, which is sufficient to avoid boundary effects but without strongly increasing the number of nodes for the sake of keeping the computational cost as low as possible. With $t_{\max}=9$ and the intercept $w_0$, each local model counts $9\times 2746+1 = 24,715$ latent Gaussian variables. 

\subsection{Bayesian inference using the integrated nested Laplace approximation}\label{sec:inla}
We represent our local dependence model, detailed in Section~\ref{sec:local}, as a latent Gaussian model with Gaussian likelihood as follows. We denote by $\mathbf{Z}=\{Z(\ss,t)\}_{t\in\mathcal{T}, \ss\in\SS}$ the  standard Gaussian observations obtained after transforming the raw anomalies using one of the marginal approaches described in Section~\ref{sec:modeling}. Precisely, using on the probability integral transform and the fitted tail parameters, we proceed with the following transformation
$$
Z(\ss,t) = \left\{
     \begin{array}{ll}
         \Phi^{-1} \left[\tilde{F}_{emp} \left\lbrace \tilde{Z}(\ss,t);\ss\right\rbrace\right] & \mbox{if } \tilde{Z}(\ss,t) \leq 0.75, \\
         \Phi^{-1} \left[ T\left\lbrace\tilde{Z}(\ss,t); \sigma_{GP}(\ss), \xi(\ss),p(\ss)\right\rbrace \right] & \mbox{if } \tilde{Z}(\ss,t) >0.75,
     \end{array}
 \right.
 $$
 where $\Phi$ is the standard Gaussian distribution function and 
 \begin{equation}
 T(z; \sigma_{GP}, \xi,p) = \left\{ 
 \begin{array}{ll}
         1- p\left( 1+ \xi \dfrac{z-0.75}{\sigma_{GP}}  \right)^{-1/\xi} & \mbox{if } \xi \neq 0, \\
         1-p \exp \left( - \dfrac{z-0.75}{\sigma_{GP}}  \right)  & \mbox{if } \xi =0,
     \end{array}
 \right. \notag 
  \end{equation}
and $\tilde{F}_{emp}( \cdot;\ss)$ the location-wise empirical distribution, to achieve an improved normalization of the very large anomalies in the dataset. Then, for each $(\ss,t)\in\SS\times\mathcal{T}$, we define the  space-time linear predictor
\begin{equation*}
    \eta(\ss,t) = \eta_0 + W(\ss,t),
\end{equation*}
according to \eqref{eq:spgaussian} with intercept $\eta_0$.
We assume a centered Gaussian prior for the intercept, such that the joint distribution of $\boldsymbol{\eta}=\{\eta(\ss,t)\}_{t\in\mathcal{T}, \ss\in\SS}$ is also Gaussian. By assuming that observations $Z(\ss,t)$ are conditionally independent given the latent Gaussian random vector $\boldsymbol{\eta}$ and a vector of hyperparameters $\boldsymbol{\theta}$, we use the following hierarchical structure:
\begin{align}\label{eq:LGMrepresentation}
	\mathbf{Z} \mid \mathbf{x},\tau_\mathbf{z} &\sim \prod_{t\in\mathcal{T}, \ss\in\SS}\varphi \left\lbrace Z(\ss,t) \mid \eta(\ss,t),\tau_\mathbf{z}^{-1}\right\rbrace,\nonumber\\
	\boldsymbol{\eta} \mid \boldsymbol{\theta} &\sim \mathcal{N}(\boldsymbol{\mu}_{\boldsymbol{\theta}}, \boldsymbol{Q}^{-1}_{\boldsymbol{\theta}}),\nonumber\\
	(\tau_\mathbf{z}, \boldsymbol{\theta})&\sim \text{p}(\tau_\mathbf{z},\boldsymbol{\theta}),
\end{align}
where $\varphi(\cdot\mid \mu,\sigma^2)$ is the Gaussian density with mean $\mu$ and variance $\sigma^2$,  $\boldsymbol{\mu}_{\boldsymbol{\theta}}$ is the mean vector and $\boldsymbol{Q}_{\boldsymbol{\theta}}$ is the precision matrix of $\boldsymbol{\eta}$.
The representation in~\eqref{eq:LGMrepresentation} corresponds to a latent Gaussian model (LGM; see, \emph{e.g.}, ~\citealp{Rue.al.2017}) where observations follow a Gaussian distribution with mean process $\eta(\ss,t)$ and a measurement error of variance $1/\tau_{\bm z}^{-1}$. The measurement error may not be necessary from a modeling stance when data are very smooth, but its presence in the model is required for implementing LGM-based inference. For smooth data, it is typically estimated to be very small, such that it becomes negligible in practice. Approximate Bayesian inference for LGMs is available through the integrated nested Laplace approximation approach (INLA; \citealp{rue2009approximate}) and its implementation, \texttt{R}-INLA~\citep{bivand2015spatial} in the \texttt{R} statistical software. With INLA,  posterior quantities of interest (that usually involve the evaluation of high-dimensional integrals) are numerically approximated using variants of the classical Laplace approximation~\citep{Tierney.Kadane.1986}. We use a recent version of INLA calling the PARDISO library for fast numerical matrix computations \citep{pardiso-6.0a,pardiso-6.0b,pardiso-6.0c,van2019new}.  INLA leverages modern numerical techniques for sparse matrices, providing fast and accurate inference for LGMs with many observations and latent variables such as in our case.

Regarding prior distributions, we choose a Gaussian distribution with precision $0.1$ for the mean parameter in~\eqref{eq:spgaussian}, which provides a moderately informative prior. The  hyperparameters $\tau_\mathbf{z}$ and $\boldsymbol{\theta} = (\kappa, \sigma_W^2,\rho)$ control the Gaussian likelihood and the latent Gaussian vector $\boldsymbol{\eta}$, respectively. Although non-informative priors are a common choice when little expert knowledge is available, we here select moderately informative prior distributions using the penalized complexity (PC) prior approach~\citep{Simpson.al.2017}. This procedure penalizes excessively complex models at a constant rate by putting an exponential prior on a distance to a simpler baseline model (\emph{e.g.}, using the baseline $1/\tau_{\bm z}=0$ for the measurement error variance), which helps to stabilize the estimation algorithm by facilitating the convergence of Laplace approximations. 
Priors then shrink model components 
toward their base models, thus preventing overfitting.
For the precision $\tau_{\bm z}$ of observations, we set an informative prior, such that the probability of observing a standard deviation larger than 0.1 is $50\%$. For the Mat\'ern covariance, we set a prior distribution where the probability that the range is smaller than 500km and that the variance larger than $0.25$ is $50\%$ in each case. Finally, an informative PC prior is chosen for the temporal auto-correlation coefficient of the AR(1) process in~\eqref{eq:spgaussian}, which reflects the strong temporal dependence: a priori, the probability of observing $|\rho|$ greater than $0.85$ is set to $50\%$.




\subsection{Generating predictive samples}
\label{sec:predsamp}

In contrast to simulation-based estimation of posterior distribution through techniques such as Markov-chain Monte-Carlo, INLA does not automatically provide a representative sample of latent variables and predictors to calculate posterior quantities of interest. Nevertheless, it is possible to obtain samples of the approximate posterior distribution efficiently: first, we sample a set of hyperparameters according to the approximation to their posterior distribution conducted by INLA; second, conditional on the hyperparameter values, we sample from the Gaussian distributions arising from the Laplace approximation~\citep{Krainski.al.2018}. 

For each local Gaussian model, we have implemented this simulation procedure to obtain $500$ samples of the latent variables of the fitted model. The simulated predictions of $Z(\ss,t)$ for unobserved locations and times are part of these samples, and we first backtransform them to the original marginal scale (if marginal transformations were applied preliminarily). Finally, the value of the cluster summary is calculated for each backtransformed posterior sample. This leads to a sample of size $500$ for each cluster summary  $X(\ss,t)$ to be predicted, and we represent predictive cdfs through the empirical cdfs of these samples.


\section{Results}~\label{sec:results}
We now discuss fitted models and assess the adequacy of our models to predict local space-time hot-spots in the anomalies by comparing goodness-of-fit diagnostics and predictive performance for the cluster summary given by the minimum. Improvements owing to explicit modeling of  marginal tails are highlighted. 

\subsection{Marginal modeling}
\begin{figure}[!t]
    \centering
    \includegraphics[width=0.95\linewidth]{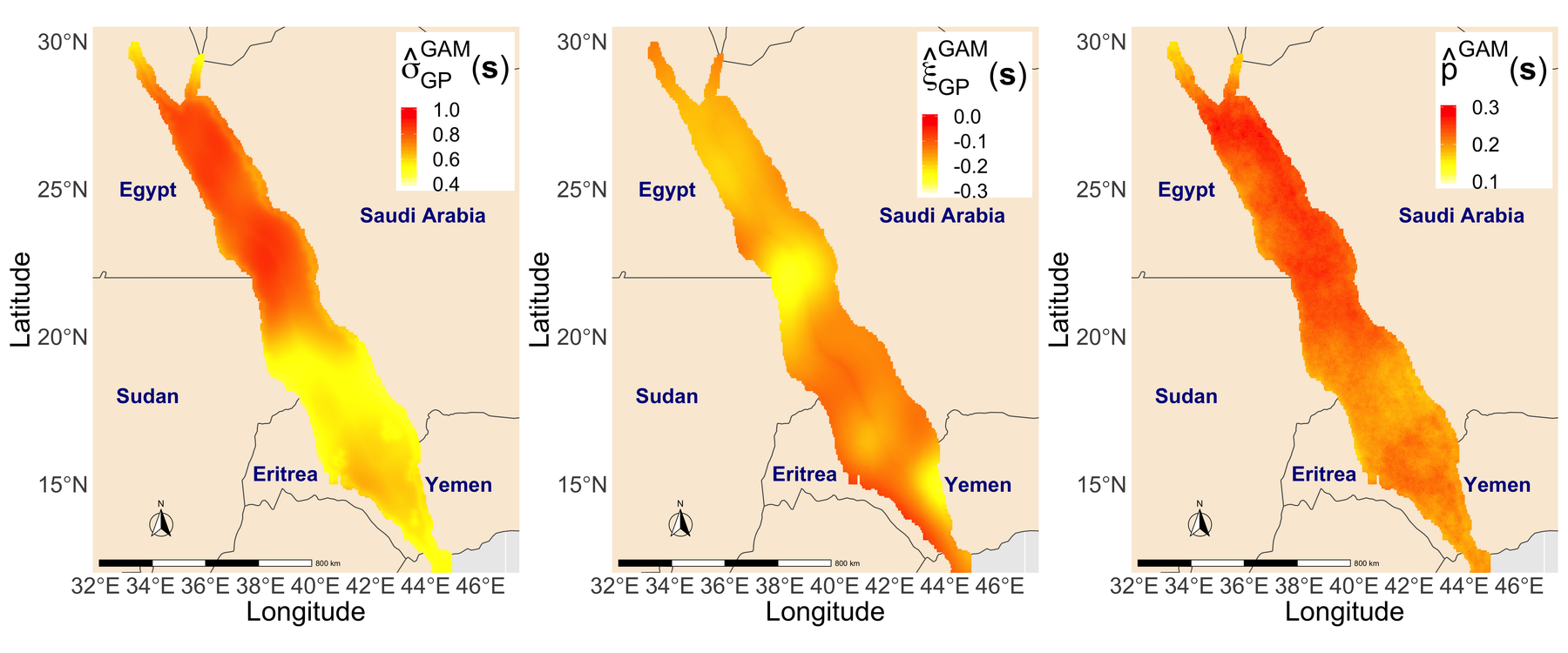}
    \includegraphics[width=0.95\linewidth]{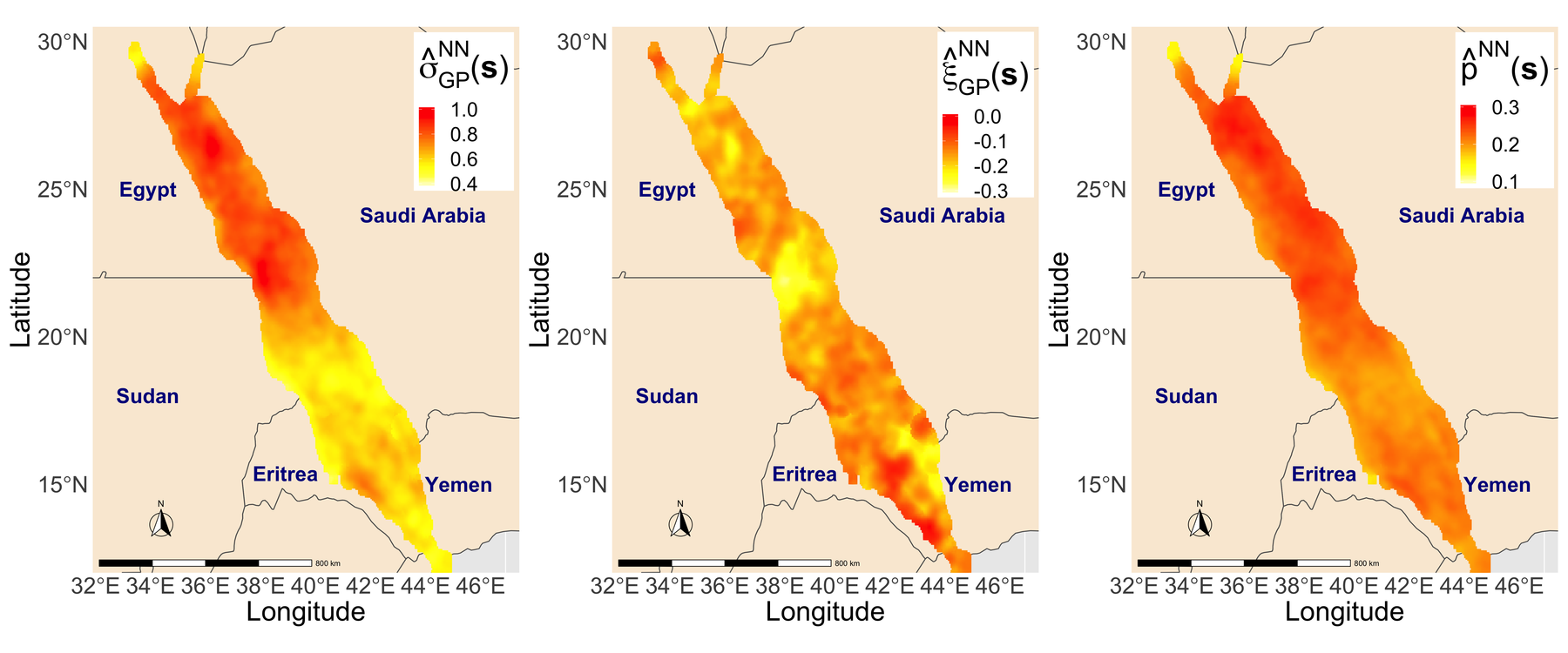}
    \includegraphics[width=0.95\linewidth]{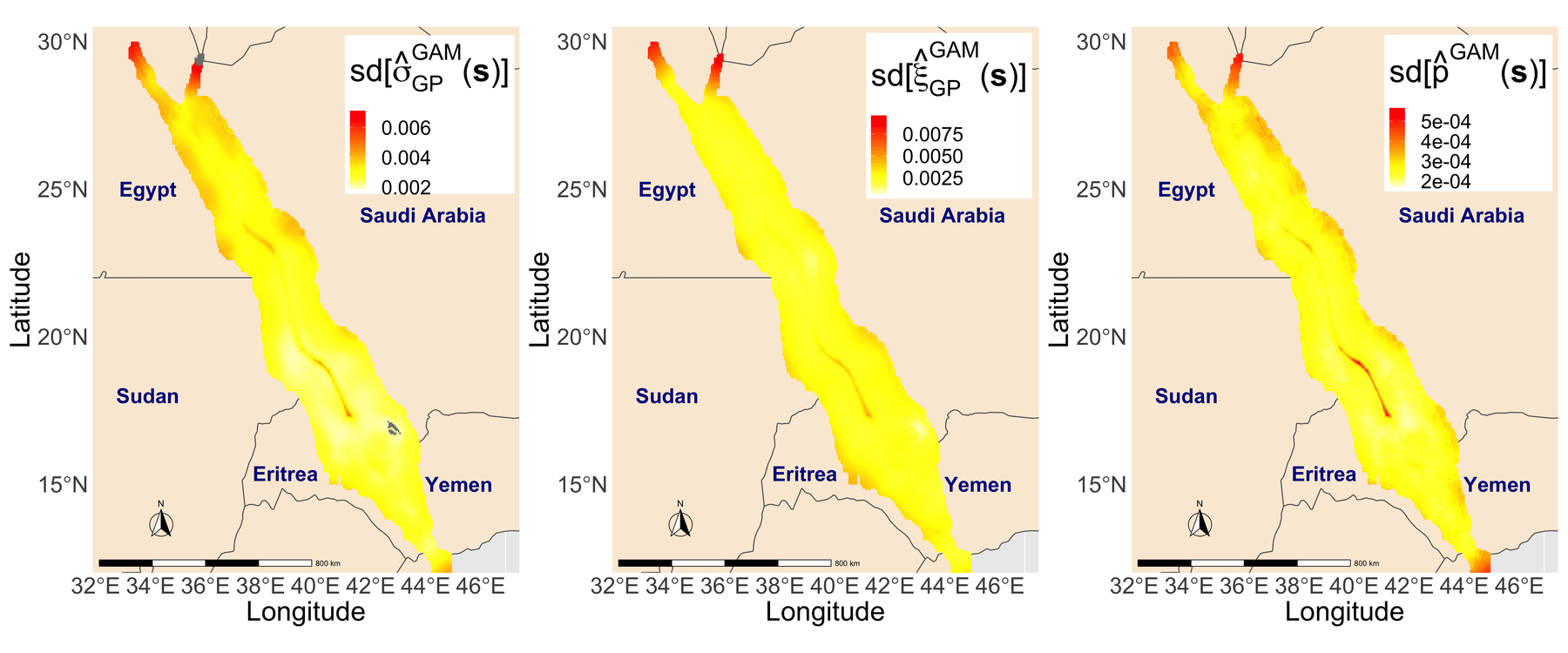}
    \caption{Generalised Pareto scale, shape, and exceedance probability  fitted to exceedances over the value 0.75 in a standard Gaussian scale using the GAM approach (top panels, standard deviations displayed in the bottom pannels) and the NN approach (middle panels).}
    \label{fig:fit_GPD_prob}
\end{figure}

The first marginal modeling step consists in capturing the temporal non-stationarity in the anomalies, \emph{i.e.}, fitting the parametric regression model in~\eqref{eq:gam_Gauss}. The fitted model parameters $\hat{\mu}(t)$ and $\hat{\sigma}(t)$ are displayed in Figure~\ref{fig:AnnualAvgYears.pdf}. The second marginal modeling step is to correct the Gaussian tail using a spatial GP distribution with either the GAM structure in ~\eqref{eq:gp_gam} or the nearest-neighbors (NN) approach detailed in Section~\ref{subsec:marginallocalmodeling}. Figure~\ref{fig:fit_GPD_prob} shows the fitted GP scale, shape, and exceedance probability across the entire region for both approaches. The fitted scale parameters display a North to South effect that is consistent with the spatial pattern in Figure~\ref{fig:TailAnomaliesLocations.pdf}. The shape is consistently negative, which implies a finite upper bound for the distribution of SST. 



\begin{figure}[!t]
    \centering
    \includegraphics[width=1\linewidth]{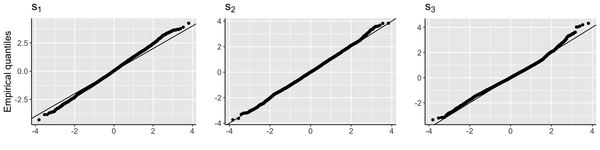} \\
    \includegraphics[width=1\linewidth]{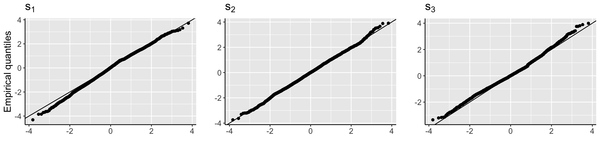} \\
    \includegraphics[width=1\linewidth]{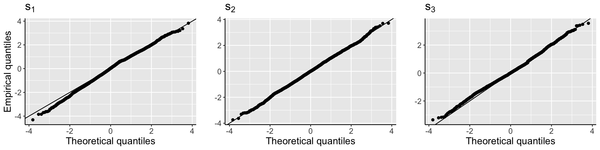}
    \caption{QQ-plots (Gaussian scale) of the normalized data $Z(\ss,t)$ with no tail-correction (top panels) and tail-correction using the GAM approach (middle panels) and the NN approach (bottom panels), at the three locations indicated in Figure~\ref{fig:AnnualAvgYears.pdf}.}
    \label{fig:qqplots}
\end{figure}

As a graphical goodness-of-fit test for the GP model, we display in Figure~\ref{fig:qqplots} the QQ-plots of the normalized data $Z(\ss,t)$ with no tail correction and with a tail correction using both the GAM and the NN approaches. Though the QQ-plots match for non-exceedances, we observe a distinct improvement in the tail region when we resort to extreme-value methods and properly model the spatial pattern in the large normalized anomalies.

\subsection{Local Gaussian dependence models}

\begin{figure}[!t]
    \centering
           \includegraphics[width=.31\linewidth]{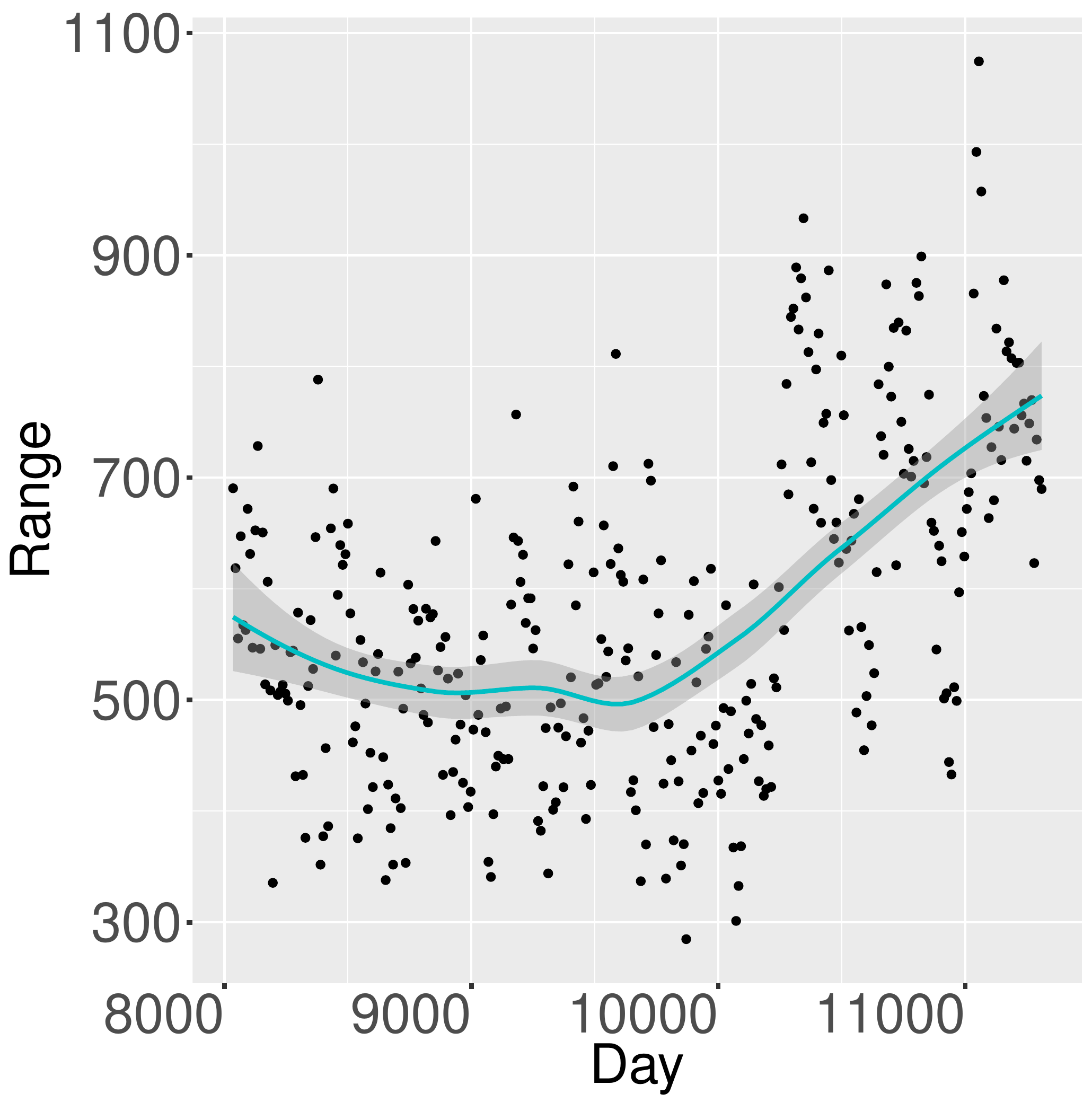}
          \includegraphics[width=.31\linewidth]{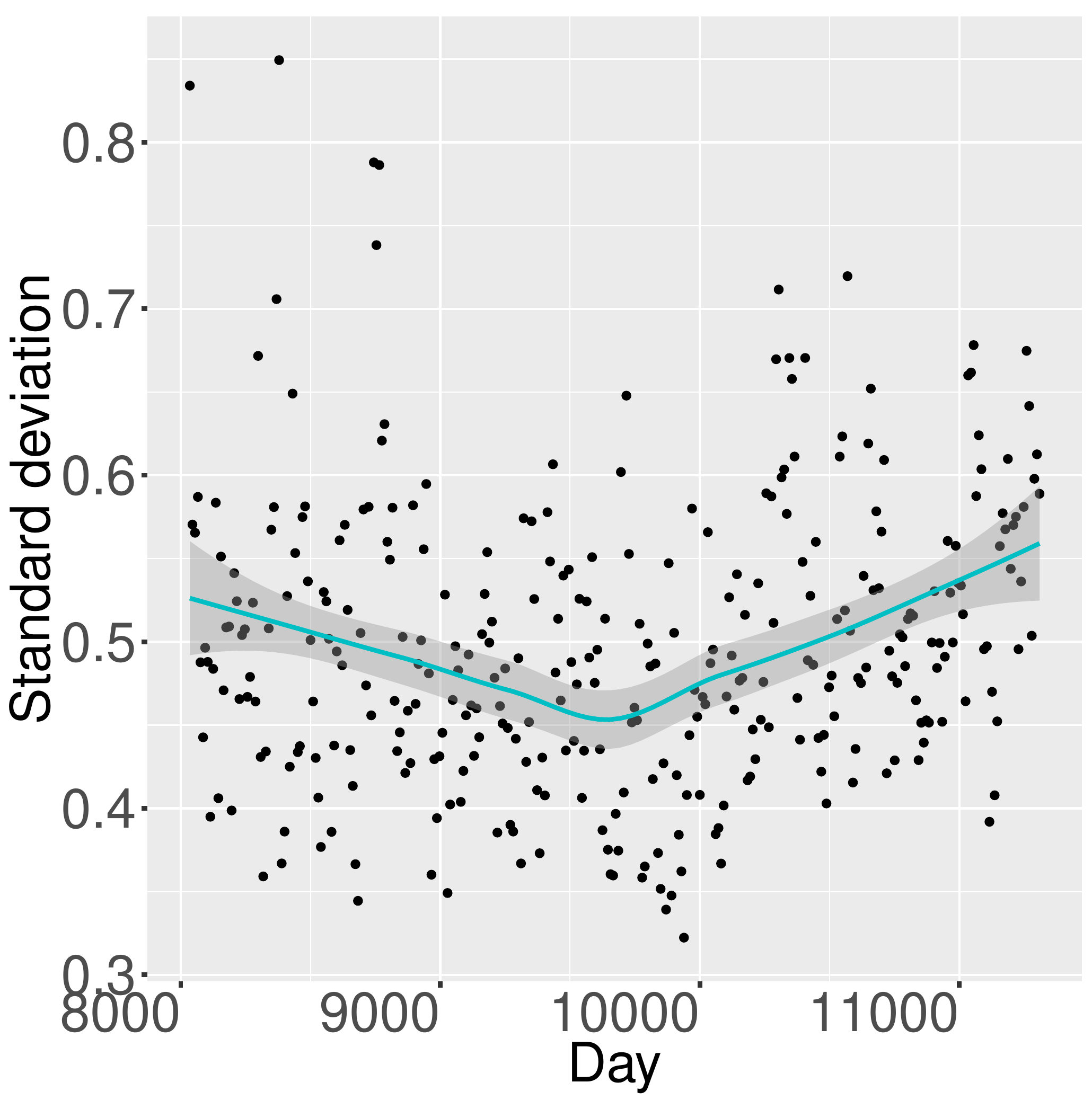}
    \includegraphics[width=.31\linewidth]{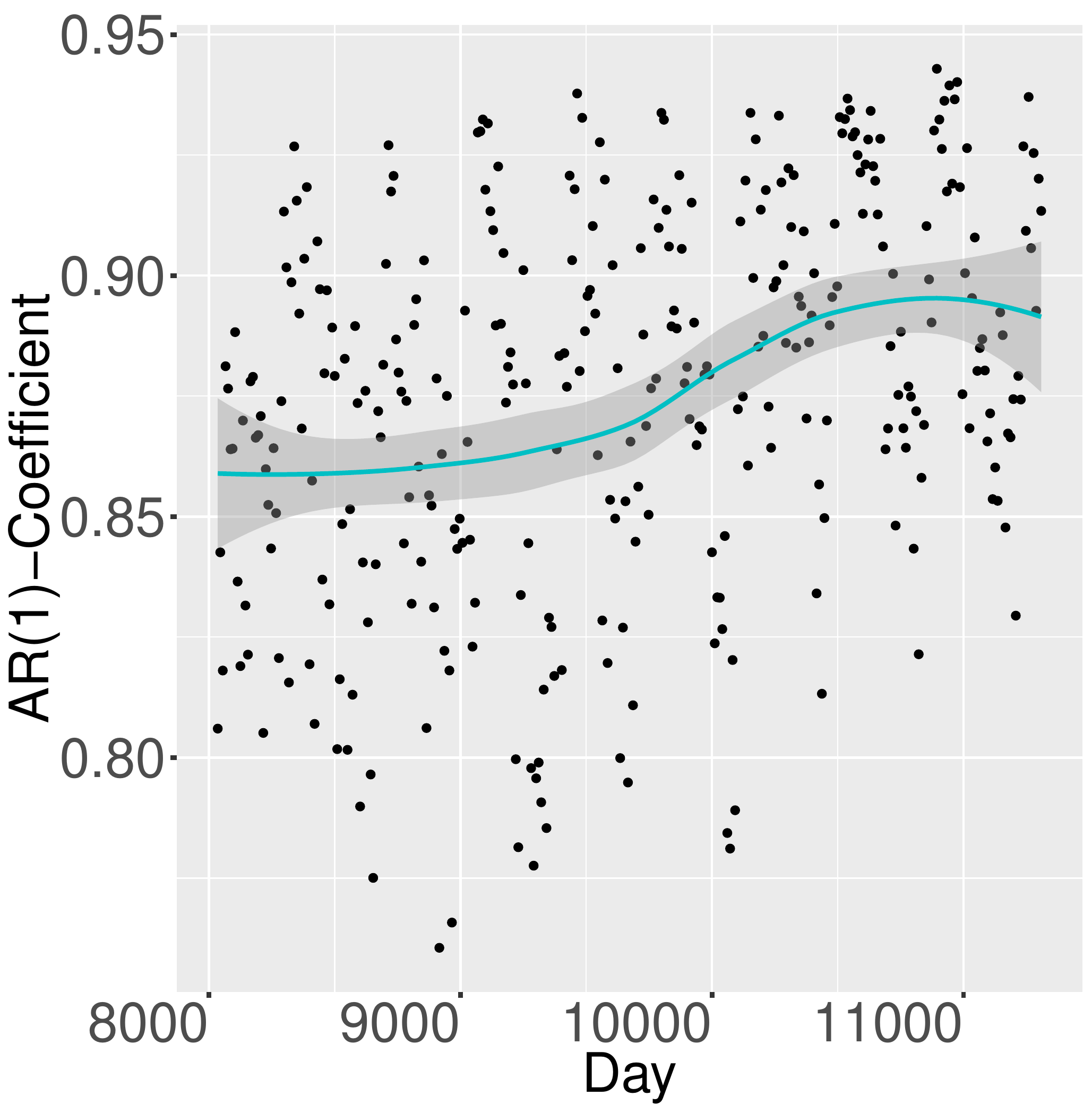} \\
               \includegraphics[width=.31\linewidth]{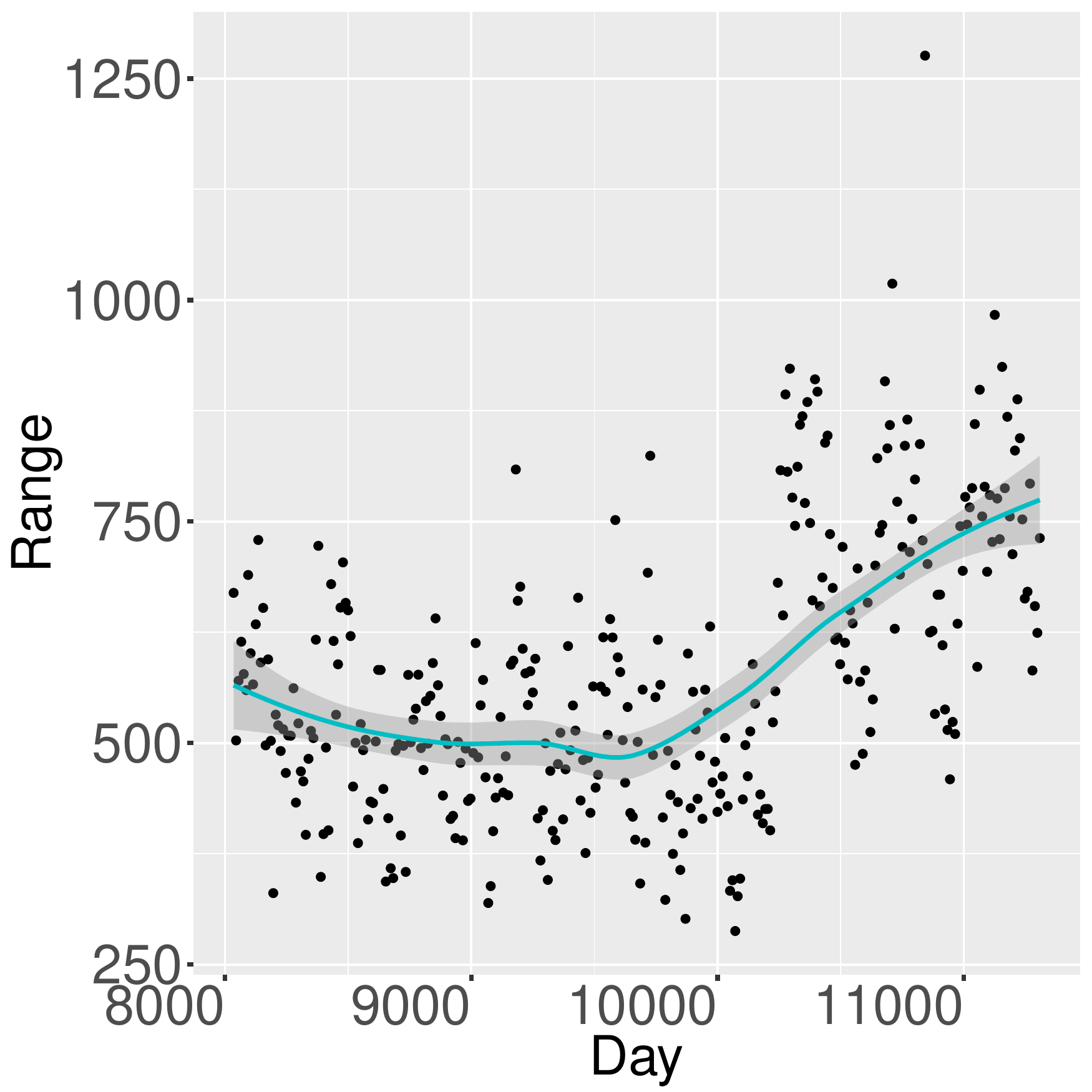}
          \includegraphics[width=.31\linewidth]{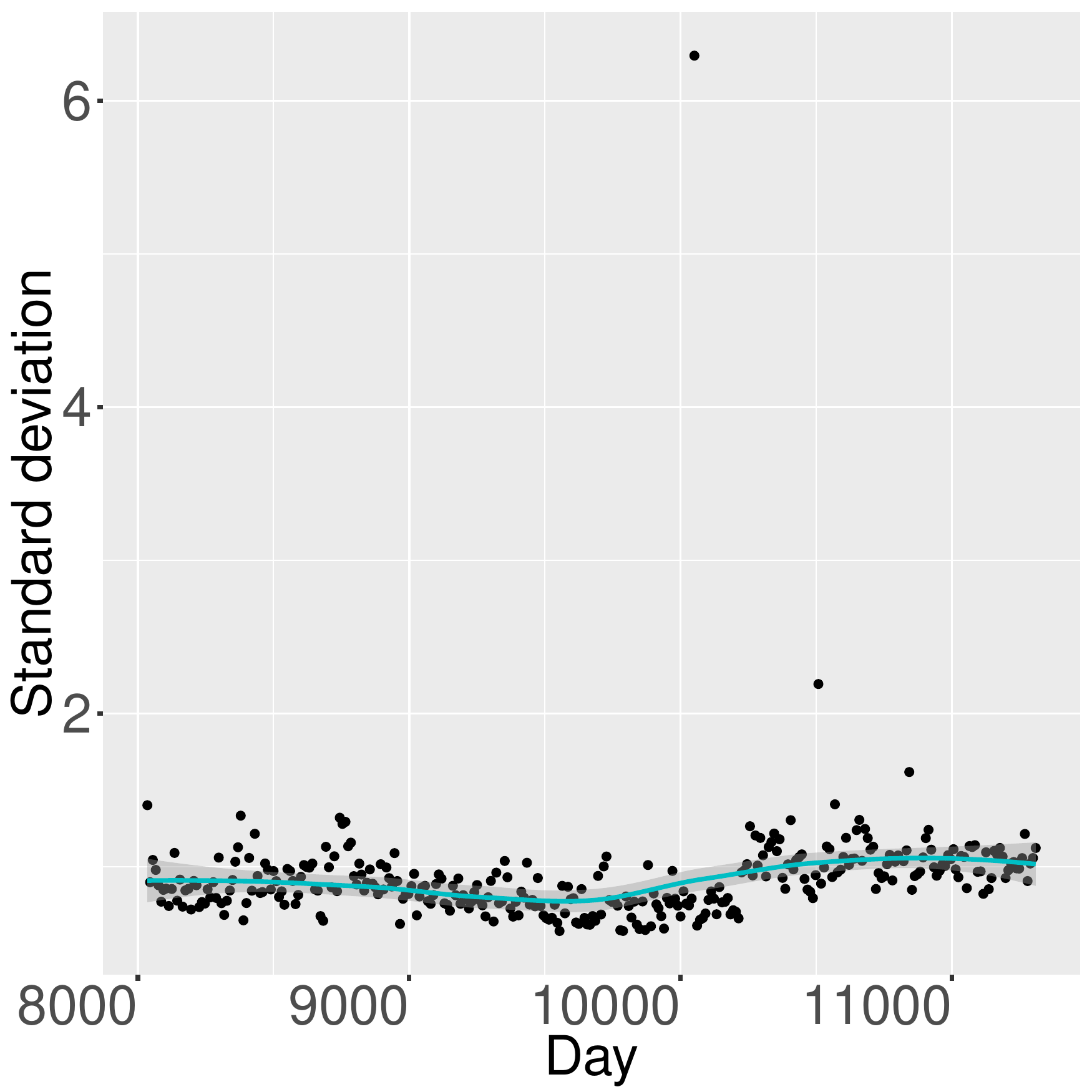}
    \includegraphics[width=.31\linewidth]{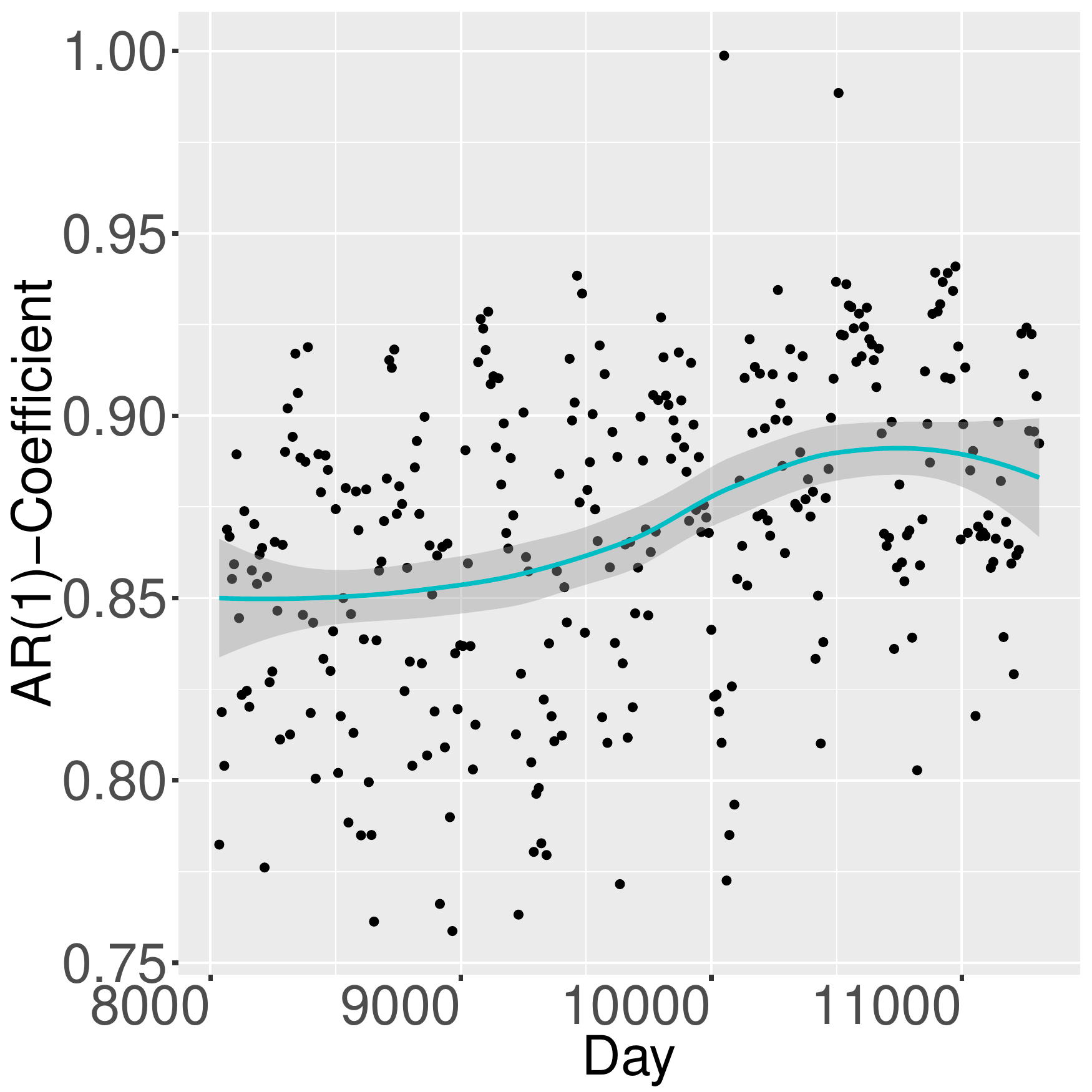}  \\
       \includegraphics[width=.31\linewidth]{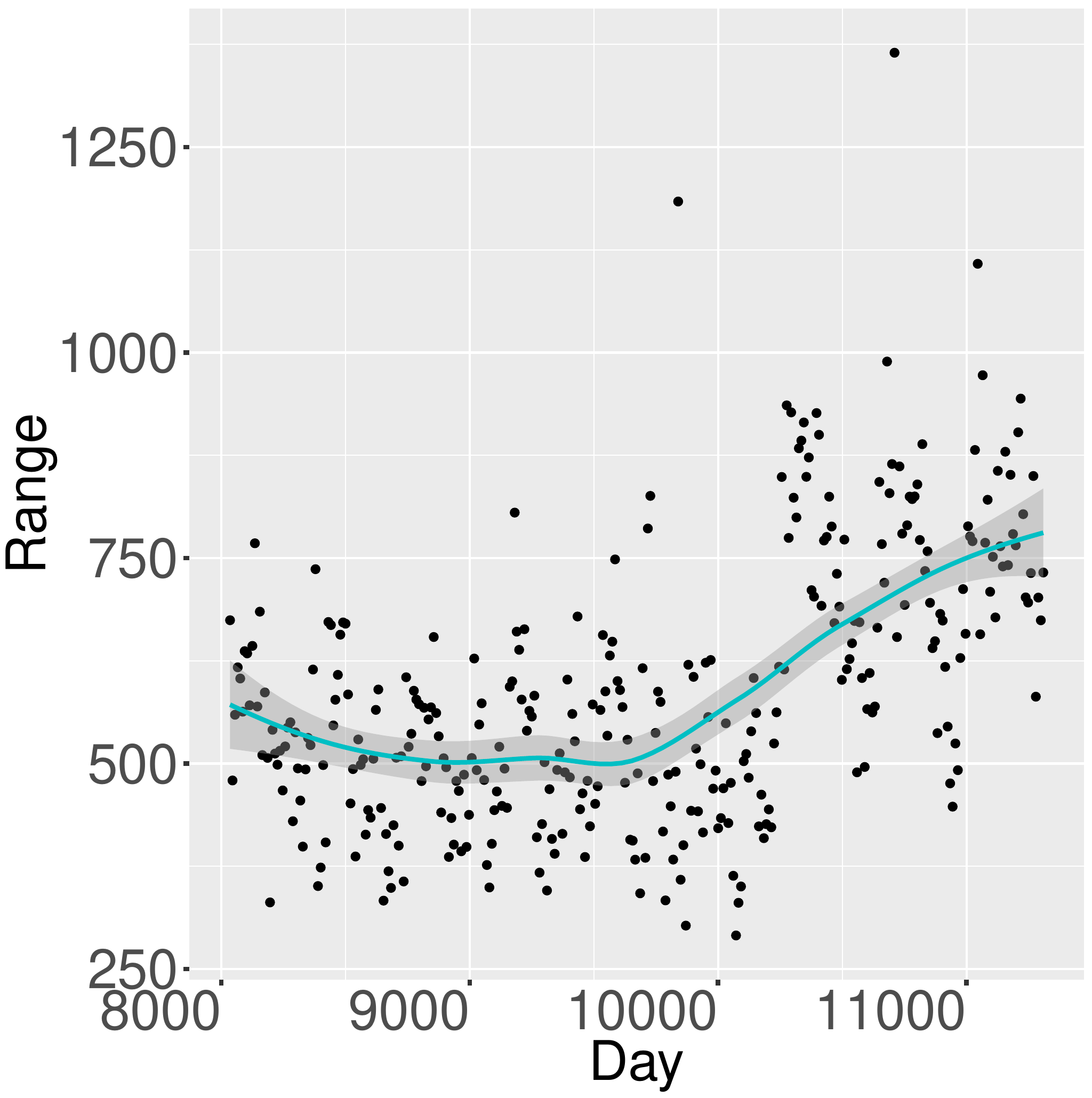}
          \includegraphics[width=.31\linewidth]{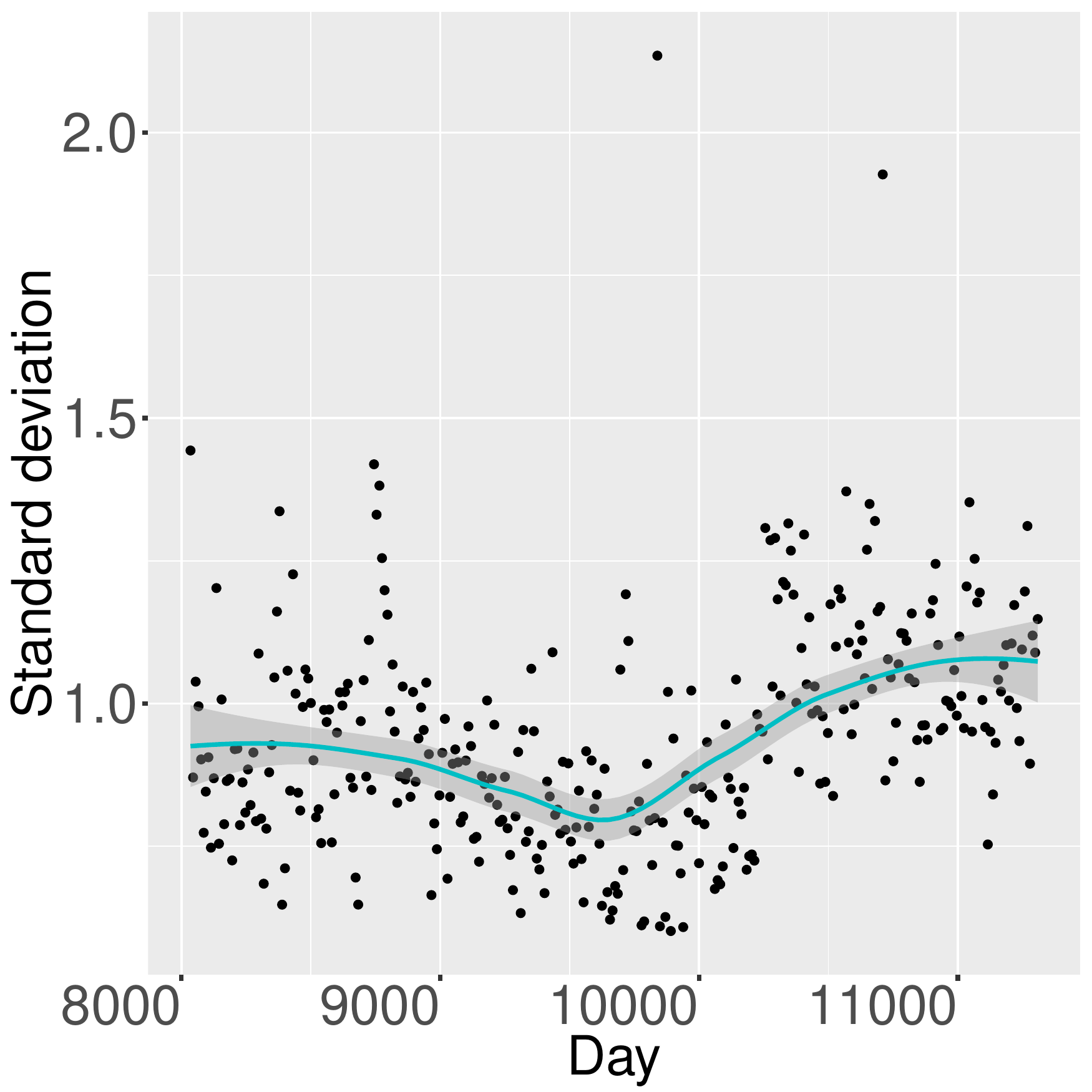}
    \includegraphics[width=.31\linewidth]{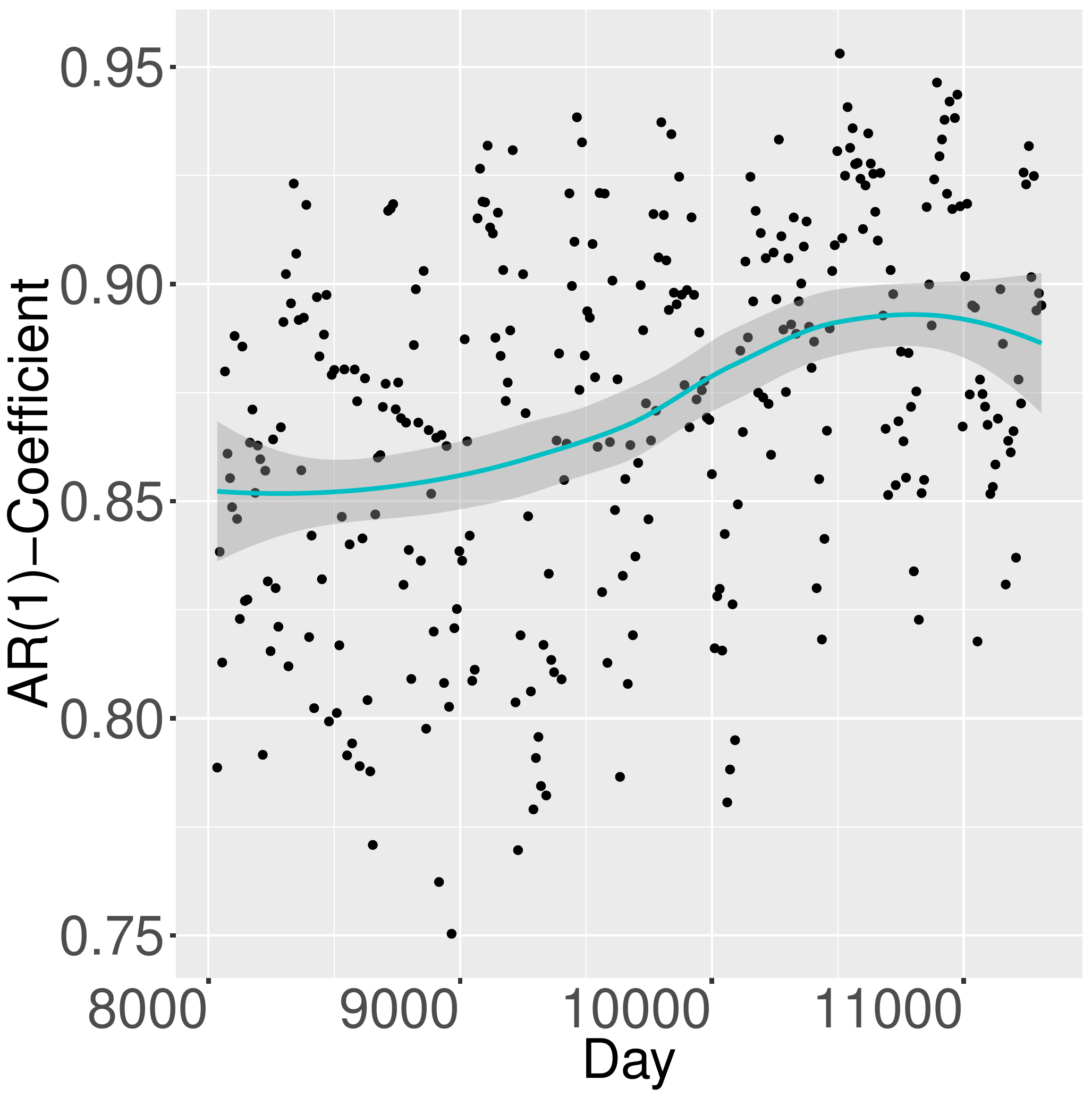}
    \caption{Posterior mean estimates of hyperparameters of the $324$ local Gaussian models, and local kernel smoothing. First row: Model without marginal transformation. Second row: marginal GAM. Third row: marginal NN model. Left: Mat\'ern range. Middle: Mat\'ern standard deviation. Right:  Temporal autoregression coefficient $\rho$.} 
    \label{fig:resultshyper}
\end{figure}

We have fitted $324$ local Gaussian models, each spanning over nine days and centered at one of the $324$ distinct validation days $t_i$. We do not report exhaustive results, but to summarize some characteristics of the fits, we show the time series of the following estimated hyperparameters in Figure ~\ref{fig:resultshyper}:  Mat\'ern range, Mat\'ern standard deviation, and temporal auto-correlation coefficient. Interestingly, some obvious trends arise over time. The three rows of the figure correspond to the three approaches to handling marginal distributions, respectively. Estimated values for the Mat\'ern range and temporal auto-correlation are very similar between the three models.
Regarding the Mat\'ern standard deviation, we obtain smaller values for Model 1 without any marginal pretransformation of data to unit variance, and one strongly outlying value arises with Model 2. All three models show the same patterns of temporal trends: ranges tend to increase towards the end of the validation period (years 2007-2015) shown in the plot, which implies that spatial dependence has increased. For the standard deviation, a pattern indicating stronger variability at the beginning and the end of the validation period arises. Finally, the temporal dependence captured in the autoregression coefficient is also increasing towards the end of the validation period, by analogy with the spatial dependence. 

\begin{figure}[!t]
    \centering
    \includegraphics[width=.31\linewidth]{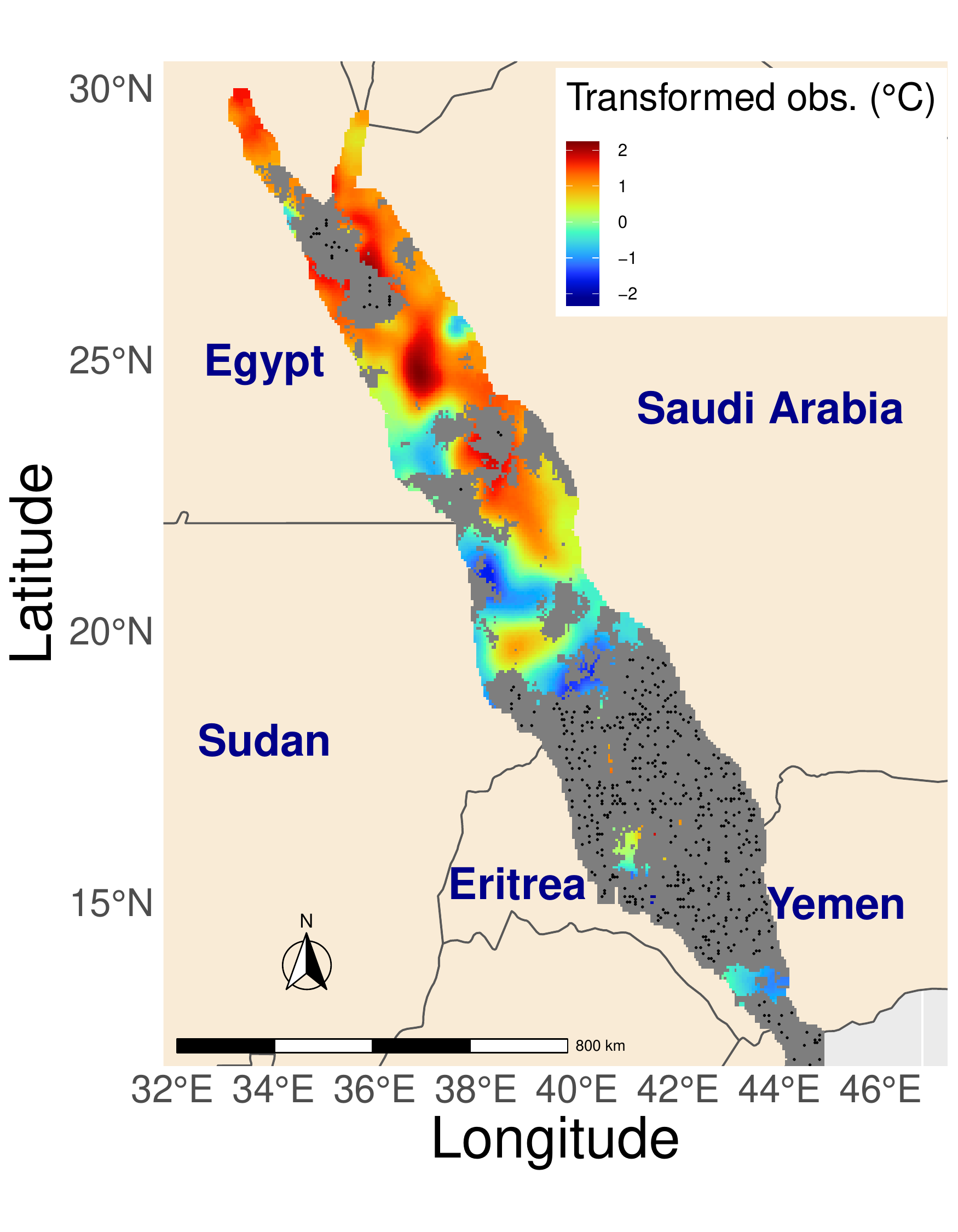} 
   \includegraphics[width=.31\linewidth]{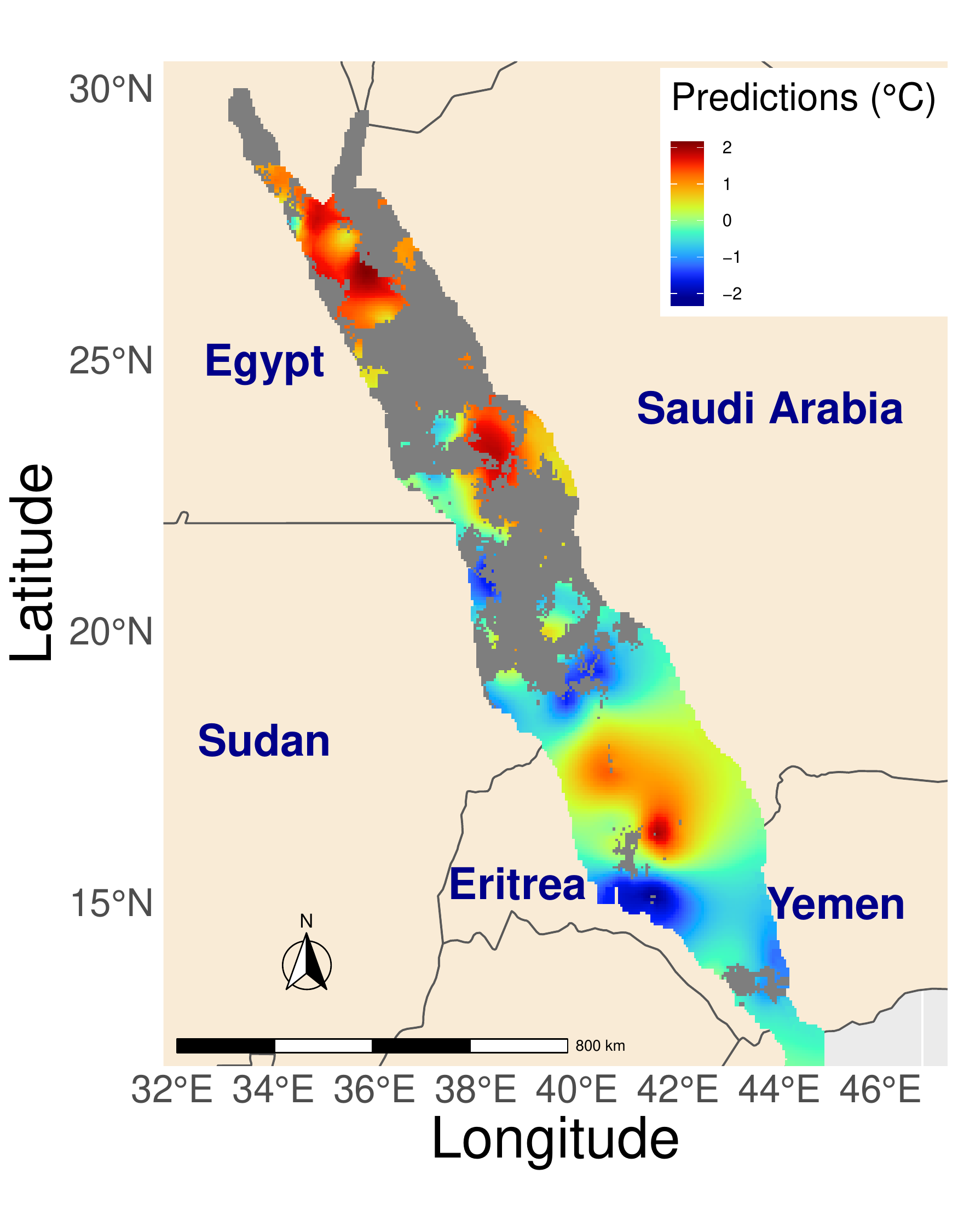}
   \includegraphics[width=.31\linewidth]{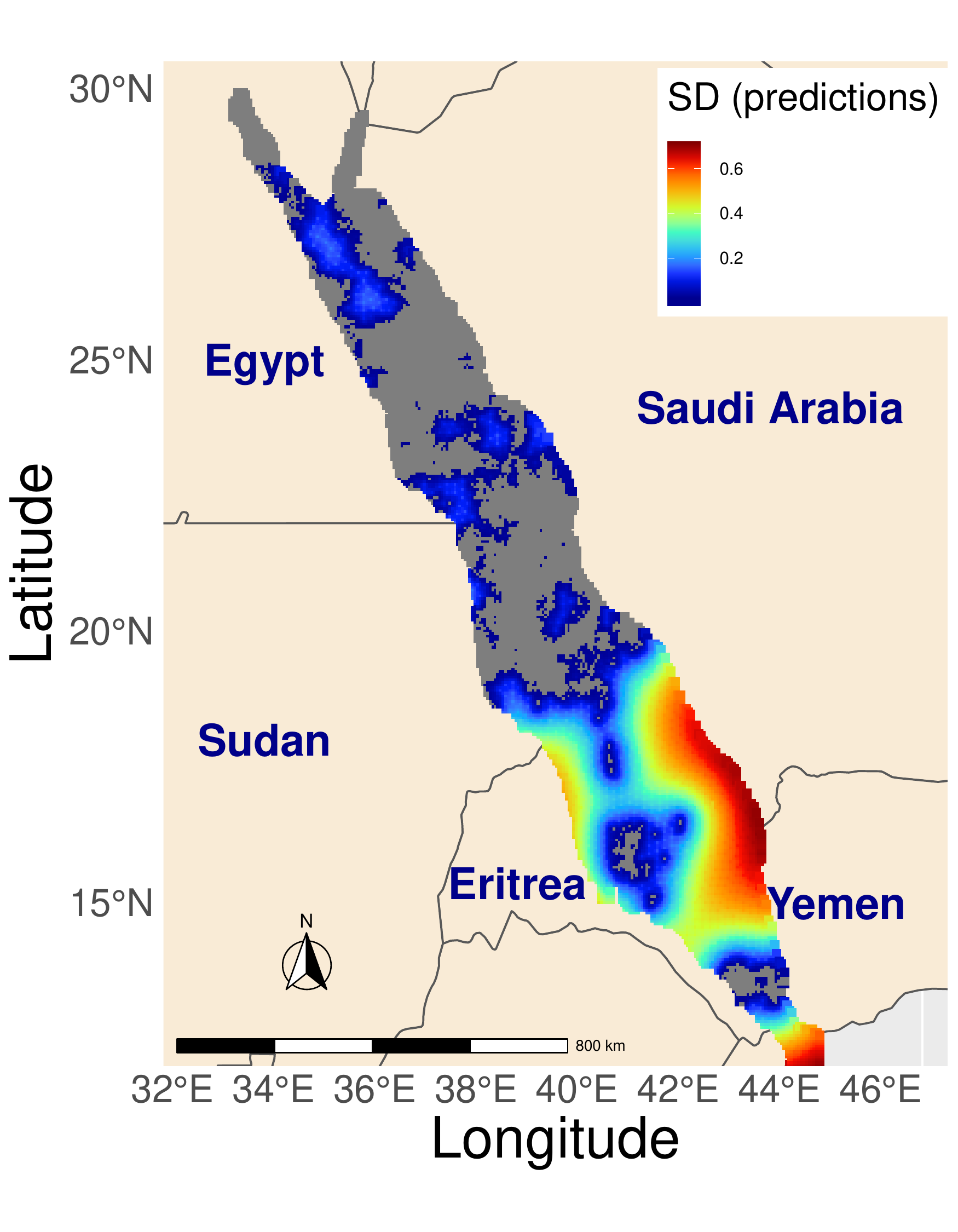}
    \includegraphics[width=.31\linewidth]{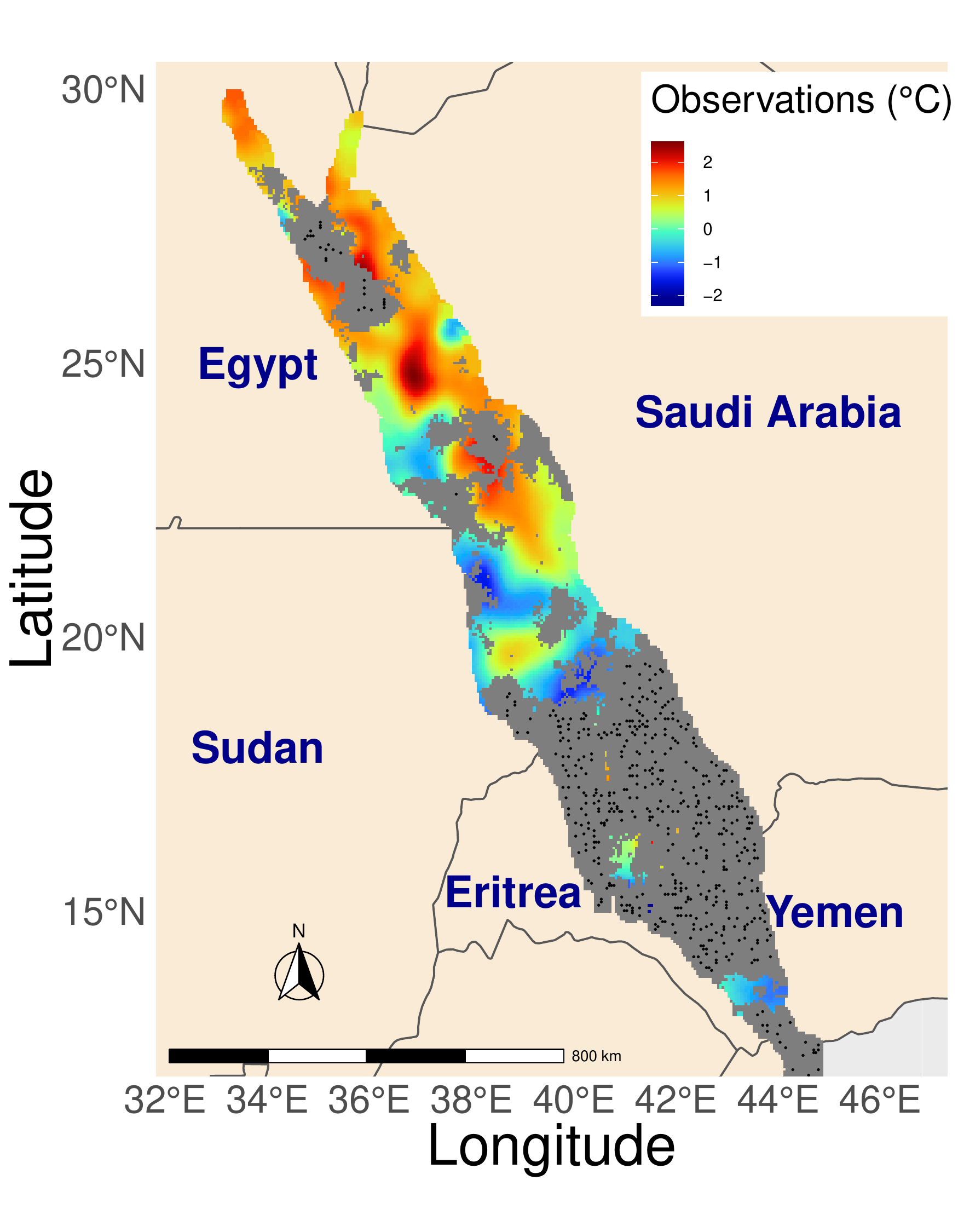} 
   \includegraphics[width=.31\linewidth]{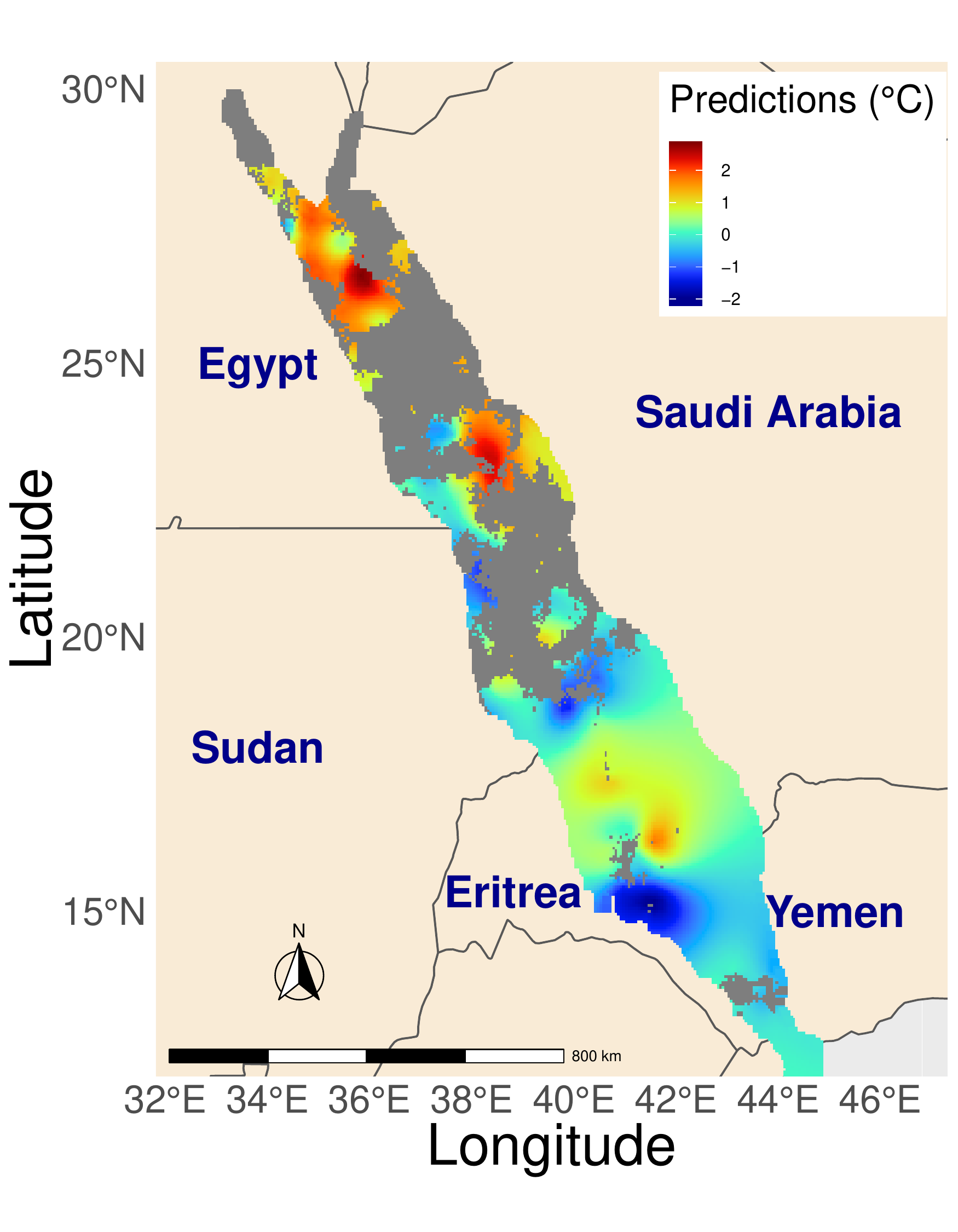}
   \includegraphics[width=.31\linewidth]{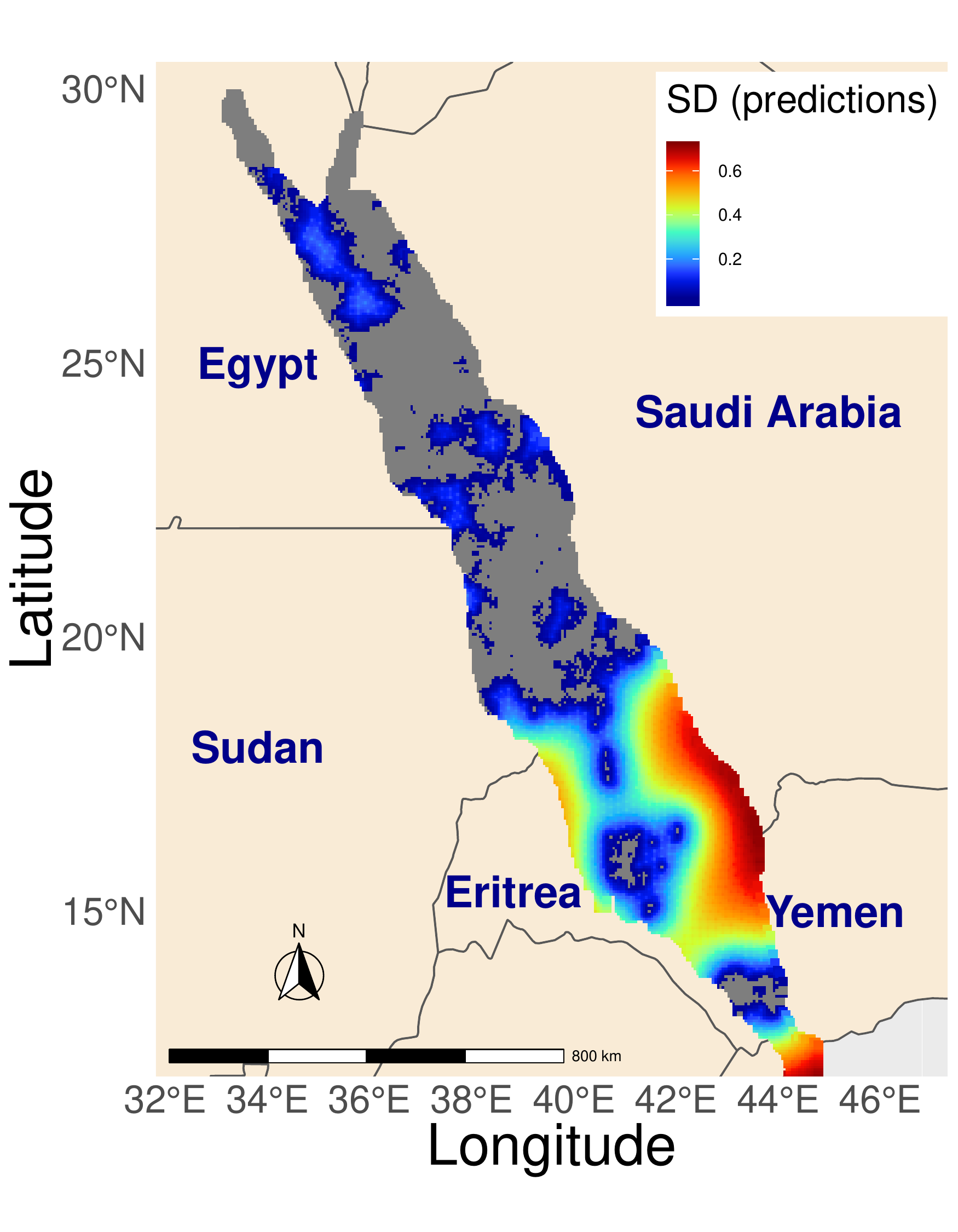}
    \caption{Example of prediction through local Gaussian model using the GAM (top) and NN (bottom) approaches. Left: Marginally transformed observations for validation day $t_i$ (in December 2009); black dots indicate validation locations. Middle: Posterior means of the predictions from the local Gaussian model. Right:  Posterior standard deviation of the predictions from the local Gaussian model.}
    \label{fig:localpred}
\end{figure}

Figure~\ref{fig:localpred} illustrates the prediction on the Gaussian scale for a validation day in December 2019. The map on the left shows the available data, pretransformed according to the nearest-neighbor approach for marginal distributions, and the validation locations. The map in the middle shows posterior means of predictions obtained through INLA. Finally, the right-hand map shows the corresponding posterior standard deviations. As expected, prediction uncertainty is lower when locations are close to observed locations but can be relatively high in the case of large gaps, such as the one in the southern part of the Red Sea.  
 
\subsection{Score evaluation}
We compare Models 1--3 by applying the tail-weighted Continuous Ranked Probability Score (\text{twCRPS}) used for ranking predictions of the EVA challenge \citep{huser2020eva}. Given an estimated predictive distribution $\hat{F}_{\ss,t}$ of the realized cluster summary $x({\ss,t})$, here defined as the minimum over space-time cylinders, the \text{twCPRS} is defined using a weight function $\omega(x)$ as
$$
\text{twCPRS}\{\hat{F}_{\ss,t}, x({s,t})\} = \int_{-\infty}^\infty \left[  \hat{F}_{\ss,t}(x) - 1 \left\lbrace x({s,t})\leq x \right\rbrace \right]^2 \omega(x)\, \mathrm{d}x, \quad \omega(x) = \Phi\{(x-a)/0.4\}, \quad a\in \mathbb{R}. 
$$
We consider the weight function of the EVA challenge with $a = 1.5$, corresponding to a Gaussian cdf that puts little weight to values below $1$. Moreover, we assess our models by evaluating the twCRPS for weight functions that put stronger weight to more extreme or less high quantiles by choosing $a=1.0$,  $a = 1.8$, and $a=-\infty$, the latter corresponding to the classical non-weighted CRPS. Table~\ref{tab:twcrps} summarizes the scores of Models 1--3 for different tail weight functions. In general, the scores of the three models are relatively close, which confirms that, locally, data are relatively smooth in space and time, such that a Gaussian dependence model performs already very well. The twCRPS scores of the GAM and nearest-neighbor (NN) marginal models are systematically improved as compared to the Gaussian model without marginal transformations, except for the classical CRPS. The performances of  GAM and NN are relatively similar, but with better twCRPS for the NN model when putting weight on predictions at high quantiles. We conjecture that this is due to some local effects that are more easily captured through local pooling than through covariate modeling, and that the resulting discrepancy seems to be amplified when predicting very high quantiles of the cluster summary.  
We finally point out that our three models outperform all models developed by other teams for the data of the EVA 2019 challenge~\citep{huser2020eva}, with their best score for $a=1.5$ being $3.67$, while a simple benchmark score for $a=1.5$ was calculated to be $7.89$. 

\begin{table}
    \centering
    \begin{tabular}{l|rrr}
         &  Gaussian  & GAM   & NN \\
\hline 
    $a=1.8$     & 0.828 & 0.825 & 0.770 \\
    $a= 1.5$    & 3.27  & 3.19 &  3.07  \\
     $a= 1.0$    & 21.7  & 21.1 &  20.9  \\
    $a = -\infty$  & 702 &  717 & 717 
    \end{tabular}
    \caption{Values of $10^4\times \text{twCRPS}$  of Models 1--3 using different tail weight functions. }
    \label{tab:twcrps}
\end{table}

\section{Discussion}~\label{sec:discussion}
We have developed a probabilistic, likelihood-based approach to fill gaps in space-time datasets with particular attention to extreme values and multivariate distributions. By estimating a generative model in a Bayesian framework, we have simulated from the posterior distribution of unobserved data points to obtain representative samples of the cluster functionals such as the minimum over space-time cylinders. 
Due to the large size of the dataset, joint estimation of margins and dependence using all data seems out of reach. 
We therefore consider a two-step approach, where we first remove trends to make data more ``Gaussian-like", especially at high quantiles. The second step consists of a Gaussian model fitted locally in space and time to the transformed data. 
Although pixel-wise observations seemed to be approximately Gaussian-distributed, the first step in the modeling procedure improved the predictive performance. 
We believe that much more substantial benefits can be expected from applying the two-step procedure to other data whose features differ more strongly from a stationary Gaussian distribution.


Our approach shows that we do not need an asymptotic model from extreme-value theory for capturing the dependence during extreme episodes. Based on our preliminary analysis showing asymptotic independence, we conjecture that our local Gaussian models are sufficiently flexible to capture spatial and temporal variations during extreme episodes. 
Moreover, the asymptotic independence in the data suggests that using non-asymptotic models such as Gaussian random fields might provide a more appropriate framework compared to the asymptotic dependence structure imposed by extreme-value models such as max-stable processes and generalized Pareto processes. For modeling spatio-temporal dependence, Gauss--Markov processes provide good tractability and scalability properties for implementing prediction with large datasets, such that our choice is also of pragmatic nature. The Red Sea SST data of OSTIA are produced using geostatistical interpolation techniques similar to kriging on original data, and they are relatively smooth in space and time due to strong dependence. This is another hint of why the Gaussian dependence models perform very well for prediction. 


Due to the artificial generation of gaps in the Red Sea data, we could validate and compare modeling approaches using the true values to be predicted. When this is not the case, external validation could be implemented through cross-validation schemes by predicting cluster functionals for space-time subsets where all observations are available.

The code written for the implementation of the approach presented in this paper is freely available from  \url{https://github.com/dcastrocamilo/EVAChallenge2019}.

\section*{Acknowledgements}
This work started when Daniela Castro-Camilo was a postdoctoral fellow at King Abdullah University of Science and Technology (KAUST). Support from the KAUST Supercomputing Laboratory and access to Shaheen is therefore gratefully acknowledged. Linda Mhalla acknowledges the financial support of the Swiss National Science Foundation. 

\baselineskip 16pt
\bibliographystyle{CUP}
\bibliography{references.bib}

\end{document}